\documentclass{aa}		
\usepackage[english]{babel}
\usepackage[utf8]{inputenc}
\usepackage{amsmath,amsfonts,amssymb,enumerate}
\usepackage{txfonts,graphics,graphicx,epstopdf,float,lscape,longtable,dcolumn,footnote,subcaption,caption,color,url,hyperref,listings}
\usepackage{rotating,afterpage}

\newenvironment{mysidewaysfigure}
	{\begin{sidewaysfigure*}\vspace*{.75\textwidth}\centering}{\end{sidewaysfigure*}}
\newcommand*\xbar[1]{{\hbox{\vbox{\hrule height 0.5pt \kern0.5ex \hbox{\kern-0.1em \ensuremath{#1} \kern-0.1em}}}}}

\usepackage{natbib}
	\bibpunct{(}{)}{;}{a}{}{,} 

\title{Appearance of dusty filaments at different viewing angles}
\author{R.-A.~Chira\inst{\ref{eso},\ref{mpia}} \and R.~Siebenmorgen\inst{\ref{eso}} \and Th.~Henning\inst{\ref{mpia}} \and J.~Kainulainen\inst{\ref{mpia}} }
\institute{European Southern Observatory, Karl-Schwarzschild-Str.\ 2, 85748 Garching, Germany\\ \email{rox.chira@gmail.com}\label{eso}
\and Max-Planck Institute for Astronomy, MPIA, K\"onigstuhl 17, 69117 Heidelberg, Germany\\\label{mpia}
}

\date{accepted on 09.05.2016, revised version \today}

\abstract
	{In recent years, there have been many studies on the omnipresence and structures of filaments in star-forming regions, as well as their role in the process of star formation. 
	These filaments are normally identified as elongated fibres across the plane of the sky. 
	But how would we detect filaments that are inclined?}
	{We aim to learn more about whether, and how, total column density or dust temperature change with respect to the line of sight.
	These variations would enable observers to use dust observations to identify and study filaments at any inclination and gain more insight into the distribution and orientations of filaments within the Galactic plane.}
	{As a first step, we perform numerical calculations on simple cylindrical models to evaluate the influence of filament geometry on the average flux density.
	After that, we apply our three-dimensional Monte-Carlo dust-radiative transfer code on two models of star-forming regions and derive maps of effective total column density and dust temperature at different viewing angles.}
	{We only see slight changes of average flux density for all cylinders we study.
	For our more complex models, we find that the effective dust temperature is not sensitive to viewing angle, while the total column density is strongly influenced, with differences exceeding an order of magnitude. 
	The variations are not injective with the viewing angle and depend on the structure of the object.
	}
	{We conclude that there is no single quantity in our analysis that can uniquely trace the inclination and three-dimensional structure of a filament based on dust observations alone.
	However, observing wide variations in total column density at a given effective dust temperature is indicative of inclined filaments. }


\begin{document}
	\maketitle

\section{Introduction}\label{intro}
	Recently, studies molecular clouds have focused more and more on their filamentary structure \citep[e.\ g.,][]{Menshchikov2010,Arzoumanian2011,Peretto2012,Andre2014}.
	These studies have revealed that filaments are omnipresent at all scales and regulate star-forming activities by carrying gas from the clouds to the cores.

	At all wavelengths, in both extinction and absorption, filaments are identified as elongated structures on the plane of the sky.
	In particular, interstellar dust is an excellent tool for tracing denser parts of the interstellar medium (ISM), as well as for determining the mass and temperature distributions of molecular clouds and filaments.
	For example, \citet{Ragan2014} and \citet{Abreu-Vicente2016} used the high extinction caused by dust grains to identify very long filaments as extended ($>$ 1$^\circ$) dark features in \textit{Spitzer} and UKIDSS observations.
	Similarly, \citet{Wang2015} found very elongated structures (meaning structures that are at least ten times longer than wide) with high contrasts and systematically lower temperatures than their surroundings using \textit{Herschel} HIGAL data.

	Filaments discovered by column density mapping data must be confirmed by line observations to ensure that they are coherent in velocity.
	Since dust emission lacks any dynamical information, the two-dimensional structures we observe in column density do not necessarily trace real three-dimensional (3D) filaments, but can also be a result of superpositions along the line of sight \citep{Juvela2012a,Smith2014b}.

	Besides, it still has to be verified whether dust extinction and emission observations reflect the same structures and whether those structures reveal all of the true filament.
	A way to disentangle this problem is to combine dust observations with radiative transfer (RT) calculations.
	\citet{Juvela2012a} and \citet{Smith2014b} applied RT on filaments formed within their (magneto-) hydrodynamical models and found that the column density profiles of resolved filaments obtained from their synthetic images were similar to observed ones, suggesting that RT effects are negligible.
	This is not the case when the filament is located further away and becomes poorly-resolved.
	Moreover, they project their filaments in such a way that their long axes are seen in the plane of the sky.
	Assuming that we can find all filaments by looking for elongated, high-density structures implies that they have preferential locations and directions within the Galactic disk.
	This is not necessarily the case and may exclude a notable fraction of filaments.

	To confirm the 3D coherence of a filament regardless of inclination, we still need line observations with highly resolved velocity spectra.
	However, dust observations by \linebreak \textit{Spitzer} (3.6 -- 160$\mu$m) and \textit{Herschel} (70 -- 500$\mu$m) and ground-based surveys as ATLASGAL (870$\mu$m) are available in high quantity and, hence, the starting point for identifying objects of interest.
	The questions we attempt to answer in this paper are:
	Would we recognise inclined filaments in dust observations?
	What are the signatures of inclined filaments in the far- \linebreak infrared (FIR) and sub-mm observations?
	And are variations along different lines of sight significant enough to be detectable?

	In Sect.\ \ref{analytic}, we examine simple, isothermal cylinders and show sight line effects based on different geometries.
	\linebreak In Sect.\ \ref{methods} we introduce the RT code that we use to model dust temperature and emission of our more complex filamentary models in Sect.\ \ref{fila}.
	The results and conclusions are summarised in Sect.\ \ref{conclusions}.

\section{Cylindrical filaments}\label{analytic}

	\begin{figure*}
		\centering
		\includegraphics[width=\textwidth]{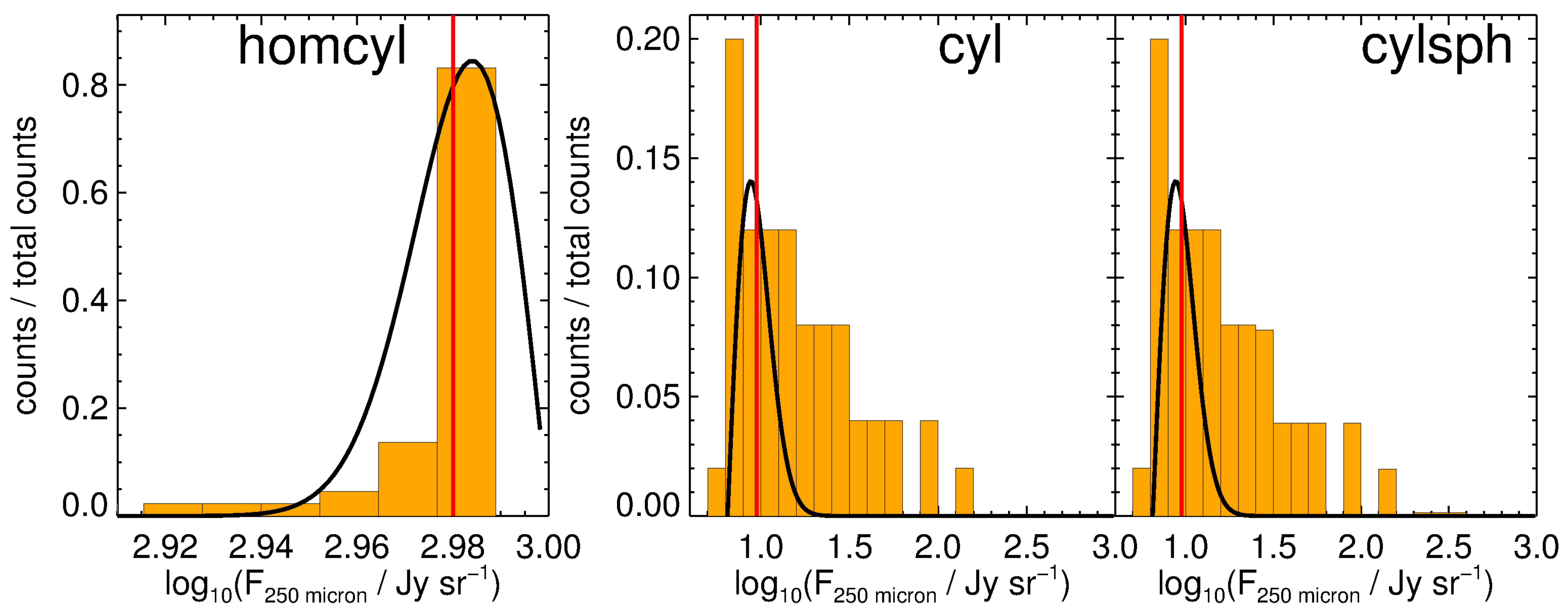}
		\caption{PDFs showing the 250 $\mu$m flux density, $F_{\rm 250 \mu m}$. 
			From left to right, the plots show the PDFs of the homcyl, cyl, and cylsph models, all at ($\theta$,$\varphi$) = (0$^\circ$,0$^\circ$) and $T_{\rm iso}$ = 10 K. 
			The black lines show the Rayleigh distributions fitted to the PDFs.
			The fitted mean values are marked with red lines.
			These plots illustrate how the flux density is concentrated towards the upper (homcyl) or lower range (cyl, cylsph) of the distributions.}
		\label{pic_cylinders_pdf}
	\end{figure*}

	\begin{figure*}
		\centering
		\begin{subfigure}[c]{.96\textwidth}
			\includegraphics[width=\textwidth]{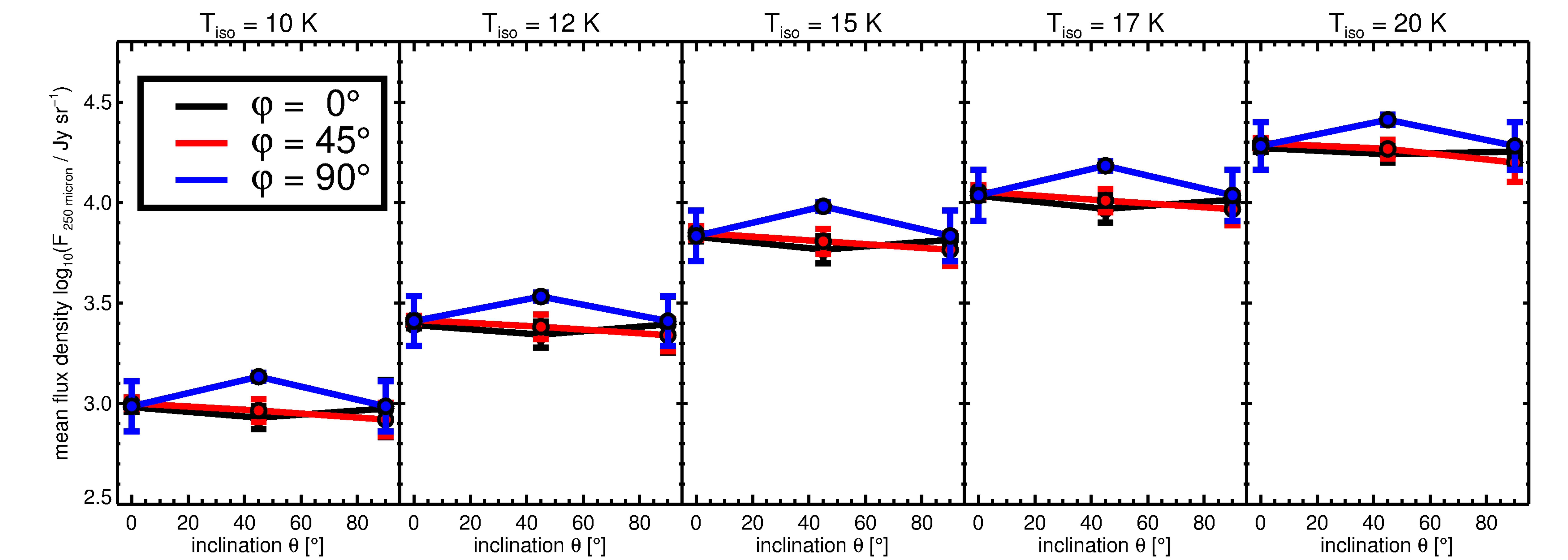}
			\caption{homcyl}
			\label{pic_homcyl_stat}
		\end{subfigure}
		
		\begin{subfigure}[c]{.96\textwidth}
			\includegraphics[width=\textwidth]{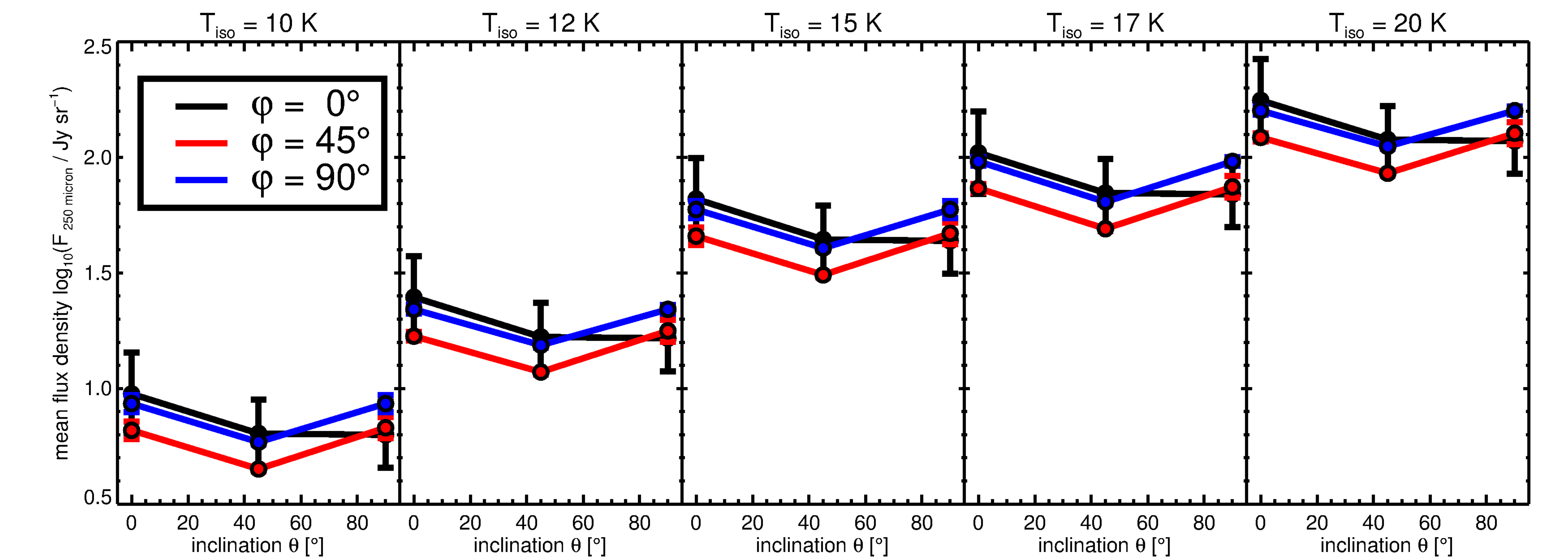}
			\caption{cyl}
			\label{pic_cyl_stat}
		\end{subfigure}
		
		\begin{subfigure}[c]{.96\textwidth}
			\includegraphics[width=\textwidth]{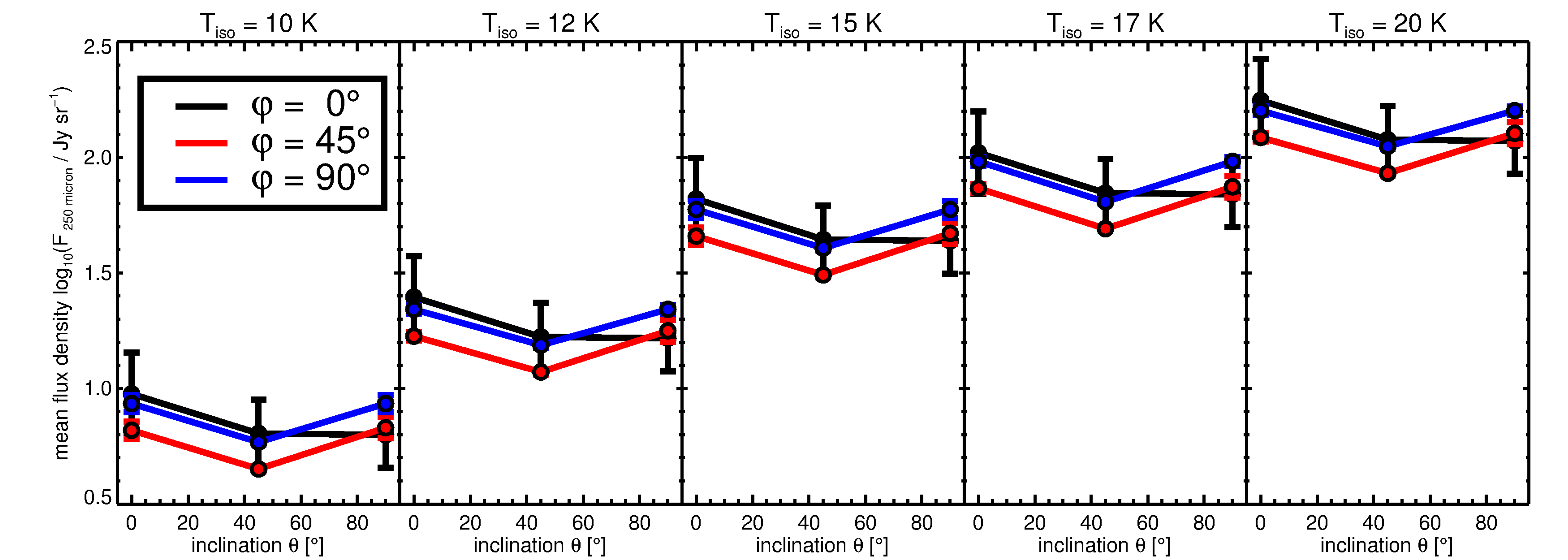}
			\caption{cylsph}
			\label{pic_cylsph_stat}
		\end{subfigure}
		
		\caption{Mean flux density, $\log_{\rm 10}$($\xbar{F}_{\rm 250 \mu m}$), of homcyl, cyl, and cylsph as function of temperature, inclination, and position angle. 
		The error bars illustrate the errors of the mean value.
		We present the results for $\theta$ and $\varphi$ at 0$^\circ$, 45$^\circ$ and 90$^\circ$. 
		The lines are interpolations that represent the true values well.
		Note that the curves for $\varphi$ = 45$^\circ$ and 90$^\circ$ may overlap in Figs.\ \ref{pic_cyl_stat} and \ref{pic_cylsph_stat}. 
		}
		\label{pic_ana_stat}
	\end{figure*}

	\begin{table}
		\begin{center}
		\begin{tabular}{l|c|c|c}
			Parameter 					& Hom.\ cylinder	& Cylinder	& Cylinder \& sphere \\ \hline
			Model ID 					& homcyl		& cyl 		& cylsph \\
			$n_{\rm x}$ / $n_{\rm y}$ / $n_{\rm z}$		& 200 / 100 / 100 	& 500 / 50 / 50	& 500 / 50 / 50 \\
			$l_{\rm cell}$ [pc]				& 0.05			& 0.05		& 0.05 \\
			$R_{flat}$ [pc]					& 0.05			& 0.03 		& 0.03 \\
			$A_{\rm V}^{\rm mean}$ [mag]			& 1947.7		& 10.2		& 11.7 \\
			$A_{\rm V}^{\rm max}$ [mag]			& 2432.0		& 63.5		& 2295.7 \\
			$A_{\rm 250 \mu m}^{\rm mean}$ [mag]		& 0.84			& 0.0044	& 0.0050 \\
			$A_{\rm 250 \mu m}^{\rm max}$ [mag]		& 1.05			& 0.027		& 0.99 \\
			\textit{cylinder}				& 			&	 	& \\
			$l_c$ [pc]					& 10.0 			& 25.0		& 25.0 \\
			$r_c$ [pc]					& 2.5 			& 1.25		& 1.25 \\
			$\rho_c$ [g cm$^{-3}$]				& 10$^{-20}$ 		& 10$^{-20}$	& 10$^{-20}$ \\
			$p_c$ 						& 0.0 			& 1.6		& 1.6 \\
			\textit{sphere}					& 			& 		& \\
			$\vec{r}_0$ [grid units]			& ---			& ---		& (0.15 $n_{\rm x}$ , $\frac{n_{\rm y}}{2}$ , $\frac{n_{\rm z}}{2}$) \\
			$r_s$ [pc]					& --- 			& --- 		& 0.2 \\
			$\rho_s$ [g cm$^{-3}$]				& --- 			& --- 		& 10$^{-19}$ \\
			$p_s$						& --- 			& --- 		& 2.0 \\
		\end{tabular}
		\end{center}
		\caption{Input parameters used for cylinders, as discussed in Sect.\ \ref{analytic}.
		Note that the numbers for mean and maximum extinction at 0.55$\mu$m, $A_{\rm V}$, and 250$\mu$m, $A_{\rm 250 \mu m}$ are valid for ($\theta$,$\varphi$) = (0$^\circ$,0$^\circ$) and may vary for other sight lines.
		}
		\label{tab_cyl_details}
	\end{table}

	\subsection{Model description}

	We start by analysing simple, isothermal models whose structure can be described analytically. 
	These models are used to explore how the mean flux density varies as a function of the viewing angle.
	Since the dust is isothermal and optically thin at FIR and submm wavelengths (see Table \ref{tab_cyl_details}), the mean column density is expected to vary similarly with the mean flux density.
	These simple models provide insight into the behaviour of the mean column density before we consider more complex models in Sect.\ \ref{fila}.

	We need to keep in mind that the flux density can only be observed relative to the fore- and background emission and is influenced by many other parameters, like dust properties, observational noise or, essentially, by the way observational data are treated (as we will see and discuss at the end of Sect.\ \ref{apply_g11}).
	This is the reason why we focus on variations in mean flux densities relative to the initial sight line only instead of analysing absolute values.

	We model filaments as cylinders \citep[since it has been done before by, for example,][]{Ostriker1964} and consider three models: \linebreak
	(a) a homogeneous, constant density cylinder (homcyl) that represents a coreless filament; 
	(b) a cylinder with a power-law radial density profile (cyl) that reflects a more realistic filament without a core; and 
	(c) a cylinder that includes a sphere with both having a power-law radial density profile (cylsph) that mimics a fragmented filament.

	Our radial density profiles are based on Plummer functions \citep{Plummer1911,Nutter2008} and are given by
	\begin{equation}
		\rho(r) = \underbrace{\frac{\rho_{\rm c}}{\left[ 1 + \left( \frac{r}{R_{\rm flat}} \right)^2 \right]^{\frac{p_{\rm c}}{2}}}}_\text{\rm cylinder component} +  \underbrace{\frac{\rho_{\rm s}}{\left[ 1 + \left( \frac{\vec{r}-\vec{r}_0}{R_{\rm flat}} \right)^2 \right]^{\frac{p_{\rm s}}{2}}}}_\text{\rm sphere component} \, ,
		\label{def_plummer}
	\end{equation}
	where $\rho_{\rm c}$ and $\rho_{\rm s}$ are the central gas densities and $p_{\rm c}$ and $p_{\rm s}$ the power-law exponents of the density profile of the cylinder and sphere component respectively, and $R_{\rm flat}$ defines the region where the density in the inner part of the cylinder is relatively constant. 
	For our calculations, we assume a gas-to-dust mass ratio $R$ = 100 and compute the dust density $\rho_{\rm i}^{dust}$ = $\rho_{\rm i} / R$.
	The quantity $\vec{r}_0$ represents the position vector of the central cell of the sphere.
	We note that the sphere component only contributes within its radius $r_{\rm s} \geq |\vec{r}-\vec{r}_0|$.
	We generate model cylinders with $n_{\rm x}$, $n_{\rm y}$, and $n_{\rm z}$ grid elements, each having the edge length $l_{\rm cell}$, in $x$, $y$, and $z$ direction, respectively.
	Table \ref{tab_cyl_details} provides the parameters we use for our calculations.
	These parameters are consistent with those \citet{Arzoumanian2011} find on average for nearby filaments.

	The dust is set to be isothermal with temperature $T_{\rm iso}$, which we vary between 10 K and 20 K.
	We rotate the models by 0$^\circ$, 45$^\circ$, and 90$^\circ$ in inclination, $\theta$, and position angle, $\varphi$, to look at the structures from different line of sights (LoSs).

	At each sight line and for each temperature, we produce flux density maps at $\lambda$ = 250 $\mu$m analytically using
	\begin{equation}
		F_{\lambda} = B_{\lambda}(T_{\rm d}) \cdot \left( 1 - e^{-\kappa_{\lambda} \, \mu_{H_2} \, m_p  \, N_{\rm tot}} \right) \, ,
		\label{def_mbnu}
	\end{equation}
	where $B_{\lambda}$ is the Planck function at the wavelength $\lambda$, $T_{\rm d}$ is the dust temperature, $\kappa_{\lambda}$ the dust opacity at $\lambda$, $\mu_{H_2}$ = 2.33 the mean molecular weight per hydrogen molecule, $m_p$ the proton mass, and $N_{\rm tot}$ the total column density along the LoS. 

	Interestingly, the scale we have chosen for $R_{\rm flat}$ is equal or smaller than $l_{\rm cell}$, which means that the flattened region of the distributions cannot be resolved.
	We tested our results by varying the ratio of $R_{\rm flat}$ / $l_{\rm cell}$ and found that our results are not affected by resolution.

	\subsection{Flux density spectral energy distribution}

	Fig.\ \ref{pic_cylinders_pdf} shows the flux density PDFs of homcyl, cyl, and cylsph at $T_{\rm iso}$ = 10 K and ($\theta$,$\varphi$) = (0$^\circ$,0$^\circ$). 
	In case of homcyl, $~$90\% of all pixels have values close to the peak flux density \linebreak (see Fig.\ \ref{pic_cylinders_pdf} \textit{left}). 
	Because of the homogeneous density distribution the flux density of homcyl is mostly peaked.
	In cyl and cylsph, only a small fraction of the models contain dust at higher density owing to the steep density gradient. 
	This means that the flux density PDFs are dominated by the low-density regime \linebreak (see Fig.\ \ref{pic_cylinders_pdf} \textit{middle} and \textit{right}).
	The sphere within cylsph contributes to the flux density PDF (within 10$^{2.2}$ -- 10$^{2.8}$ Jy sr$^{-1}$), but makes up only a small fraction of the whole volume.

	To quantify the flux density PDFs, we fit Rayleigh distributions to them which are given by
	\begin{equation}
		f(x) = A \cdot \frac{x - x_{\rm 0}}{\sigma^2} \cdot e^{- \frac{\left(x - x_{\rm 0}\right)^2}{2 \sigma^2} } ,
		\label{def_rayleigh}
	\end{equation}
	\noindent where A is a scaling factor for the amplitude, $x_{\rm 0}$ is a parameter shifting the distribution along x and $\sigma$ is the standard deviation.
	The mean value $\bar{x}$, which is equivalent to the position of the distribution peak, is given by $x_{\rm 0} + \rm{sgn}(A) \sqrt{\frac{\pi}{2}} \sigma$. 
	Since the standard deviation is small compared to $x_{\rm 0}$, the mean value is dominated by $x_{\rm 0}$.
	We still use the distribution mean value for our analysis.
	In Fig.\ \ref{pic_ana_stat} we show the mean values of the flux density PDFs, $\xbar{F}_{\rm 250 \mu m}$, as function of dust temperature, inclination, and position angle.

	We note that from now on we discuss only the behaviour of the mean values.
	The mean value is certainly the most favourable statistical parameter to compare with observational studies since it is commonly used, reflects the overall characteristics of the object, and is less affected by resolution effects.

	Alternatively, we use the maximal flux density of the distribution or its standard deviation.
	The advantage of looking for the areas of maximal flux density is that it is the most straightforward way to identify the most dense or warmest regions within a filament (for instance, the maximal flux density would be the best parameter in differentiating between cyl and cylsph).

	The standard deviation $\sigma$ (see Equ.\ (\ref{def_rayleigh})) is more promising than the maximal flux value to describe the overall properties of the models.
	We find that there are variations in $\sigma$ for sight lines along $\varphi$ = 0$^\circ$ and $\varphi >$ 0$^\circ$ (meaning both \linebreak $\varphi$ = 45$^\circ$ and 90$^\circ$).
	These variations are less revealing since they are independent of the dust temperature.
	Thus, contrary to the mean value, we would miss insights in the evolution of the filament.
	Furthermore, the variations are not uniform.
	While $\sigma$ increases from 0.01 Jy sr$^{-1}$ along $\varphi$ = 0$^\circ$ to about 0.13 Jy sr$^{-1}$ along $\varphi >$ 0$^\circ$ for homcyl, it decreases for cyl and cylsph from \linebreak 0.13 Jy sr$^{-1}$ along $\varphi$ = 0$^\circ$ to about 0.05 Jy sr$^{-1}$ along $\varphi >$ 0$^\circ$.
	This means that the changes are indeed significant in numbers, but they also depend on how the matter is distributed.
	Additionally, a sampling of $\sigma$ requires the filament to be well-resolved, which might become a problem the fainter and further it is located from us.

	For our test cases, we see that the mean flux increases with increasing temperature.
	For a given temperature, the average variations are on the order of 0.2 dex, which corresponds to a factor of about 1.58.
	This is in agreement with \citet{Arzoumanian2011} who predicted that their observed column densities are on average overestimated by a factor of \linebreak $<\cos(\theta)^{-1}> \, = \pi/2 \approx$ 1.57 owing to the unknown inclinations.

	In homcyl, we see that the mean flux density is mostly constant, with small variations within 0.2 dex.
	In the case of cyl and cylsph (Figs.\ \ref{pic_cyl_stat} and \ref{pic_cylsph_stat}), the variations are slightly larger, but not on the order of 0.3 dex (corresponding to a factor of 2.0).
	In general, the mean flux density decreases with increasing inclination.
	This is because the area within the flux density is amplified, due to larger amounts of dust along the LoS, because smaller and statistically less significant.
	Thus, the PDFs are even more dominated by the more diffuse dust, decreasing the value of $\xbar{F}_{\rm 250 \mu m}$.

	When comparing the results of cyl and cylsph, we see no contribution of the sphere to $\xbar{F}_{\rm 250 \mu m}$. 
	This means that the sphere is too small compared to the whole cylinder that dominates the total flux.
	We performed additional tests and found that increasing the density by several order of magnitude of the sphere alone does not have any impact.
	If we insert more spheres such that the spheres contain at least 10\% of the total mass of the models, we see notable fluctuations in the mean flux density.
	This means that, as long as a core is small in size and mass compared to the filament, it can only be studied directly by removing the filamentary background. 
	For example, \citet{Montier2010} filtered the local background from \textit{Planck}-HFI maps to reduce the influence of large-scale structures on the source detection and build up their \textit{Planck} catalogue of cold cores.
	This is comparable to using the maximum flux density to identify objects of interest instead of the mean flux density.

	In summary, we predict little to no variations in the mean flux density PDFs for elongated cylindrical structures based on their geometry.
	They are not significant enough to allow conclusions on viewing angles based on dust observations alone.
	We neither see significant changes in the PDF distributions when inserting a core-like sphere into our cylinder. 
	This implies that as long as a single core does not contain a significant fraction of the mass compared to its surrounding filament it does not influence the average properties of the whole filament.

\section{3D Monte Carlo dust radiative transfer code}\label{methods}

	We use a three-dimensional (3D) vectorised Monte Carlo (MC) code by \citet{Heymann2012}, which is based on the original radiative transfer code of \citet{Kruegel2008} to model realistic filaments.
	The code is specialised for radiative transfer (RT) of dust and computes self-consistently the temperatures of the dust species.
	The program keeps track of all scattering and absorption/re-emission processes, and produces flux density maps at requested viewing angles.

	The code includes different methods for optimising performance such as the iteration-free method by \citet{Bjorkman2001}.
	If a photon package is absorbed, this method ensures that the whole energy of this package is used to heat the dust with respect to previous absorption events. 
	After the dust temperature of the cell is adjusted, the energy is re-emitted as a new photon package (see \citealt{Bjorkman2001} for more details, as well as \citealt{Baes2005,Kruegel2008}).

	We extend MC to utilise arbitrary density structures that are heated isotropically by an external radiation field (ERF) by photon packages that are launched from the outer walls.
	The launching points are computed for each photon package individually by setting them randomly onto the surface of the data cube.
	The initial direction of each photon package is calculated such that the resulting external radiation field is isotropic (see also Appendix \ref{app_mu}).

	Contrary to point sources, the luminosity of an ERF depends on both the total flux density, 
	\begin{equation}
		F = \int F_\nu^{\rm ext} d\nu \, ,
		\label{def_tot_flux}
	\end{equation}
	\noindent where $F_\nu^{\rm ext}$ represents the flux of the ERF at the frequency $\nu$, and the surface area of the cube, $A_{\rm tot}$, that emits the flux.
	For controlling the luminosity, we introduce the scaling factor $w$.
	The total luminosity, $L_{ext}$, is then given by:
	\begin{equation}
		L_{\rm ext} = w \cdot \, \pi \, A_{\rm tot} \cdot F \, .
		\label{equ_def_lisrf}
	\end{equation}
	\noindent In this way, any ERF can be used.

	In Appendix \ref{app_bench} we provide a detailed description of the benchmark runs we have performed to verify the correctness of our implementations.

	\begin{table}
		\renewcommand{\arraystretch}{1.5}
		\begin{center}
		\begin{tabular}{l|c|c}
			Parameter & $\rho$ Ophiuchi & G11.11 \textit{Snake} \\ \hline
			Cube size $n_{\rm x}\, \times \,n_{\rm y}\, \times \,n_{\rm z}$	& 370 $\times$ 242 $\times$ 243	& 1181 $\times$ 581 $\times$ 581 \\
			Edge length $l_{\rm cell}$ [pc]			& 0.0473			& 0.02 \\
			ISRF strength $w$ 				& 1.0				& 1.0 \\
			Total gas mass [$M_\odot$]			& 8100.0			& 49745.0 \\
			Maximal $A_{\rm V}^{\rm z-axis}$ [mag]		& 40.4				& 70.0 \\
		\end{tabular}
		\end{center}
		\caption{Input parameters used for filamentary models, as discussed in Sect.\ \ref{fila}.
		}
		\label{tab_models_details}
		\renewcommand{\arraystretch}{0.66}
	\end{table}

	\begin{table}
		\renewcommand{\arraystretch}{1.3}
		\begin{center}
		\begin{tabular}{l|c|c|c|c}
			Quantity & Mean & Minimum & Median & Maximum \\ \hline
			\multicolumn{5}{l}{$\rho$ Ophiuchi} \\ \hline
			$\rho_{\rm gas}$\footnotemark[1] [10$^{-22}$ g cm$^{-3}$]	& 2.7	& 0.05	& 1.4	& 835.7 \\
			$N_{\rm tot}^{\rm real}$ [10$^{21}$ cm$^{-2}$]	& 5.2	& 0.001	& 3.4	& 220.0 \\
			$N_{\rm tot}^{\rm eff}$ [10$^{21}$ cm$^{-2}$]	& 13.0	& 0.001	& 6.0	& 610.0 \\
			$T_{\rm d}^{\rm eff}$ [K]			& 12.9	& 9.7	& 12.6	& 20.0 \\
			$\xbar{F}_{\rm 250 \mu m}$ [Jy sr$^{-1}$] 	& 12.3	& 6.7	& 13.5	& 19.9 \\ \hline
			\multicolumn{5}{l}{G11.11 \textit{Snake}} \\ \hline
			$\rho_{\rm gas}$\footnotemark[1] [10$^{-22}$ g cm$^{-3}$]	& 11.0	& 0.07	& 5.3	& 1858.5 \\
			$N_{\rm tot}^{\rm real}$ [10$^{21}$ cm$^{-2}$] 	& 20.0	& 0.001	& 14.0	& 690.0 \\
			$N_{\rm tot}^{\rm eff}$ [10$^{21}$ cm$^{-2}$] 	& 6.9	& 0.001	& 3.3	& 500.0 \\
			$T_{\rm d}^{\rm eff}$ [K] 			& 13.1	& 8.6	& 12.6	& 23.9 \\
			$\xbar{F}_{\rm 250 \mu m}$ [Jy sr$^{-1}$] 	& 6.6	& 2.4	& 7.6	& 10.9 
		\end{tabular}
		\renewcommand{\arraystretch}{1.0/0.75}
		\end{center}
		\caption{Statistical summary of derived parameters based on the $\rho$ Ophiuchi and the \textit{Snake} models.
			Shown are the input total column density, $N_{\rm tot}^{\rm real}$, as derived directly from the input volume density models, the effective total column density, $N_{\rm tot}^{\rm eff}$, representing the column density and the effective dust temperature, $T_{\rm d}^{\rm eff}$, both derived from SED fitting, and the mean flux density at 250 $\mu$m based on our synthetic images.
			The given numbers are based on the values at all viewing angles.
			We note that we show the arithmetic mean values here, which do not necessarily need to match the mean values of the Rayleigh distributions that we discuss in Sects.\ \ref{apply_rhooph} and \ref{apply_g11}.
		}
		\label{tab_models_derived}
	\end{table}

\section{Filaments as seen at different viewing angles}\label{fila}

	\begin{figure*}
		\begin{subfigure}[c]{0.485\textwidth}
			\includegraphics[width=\textwidth]{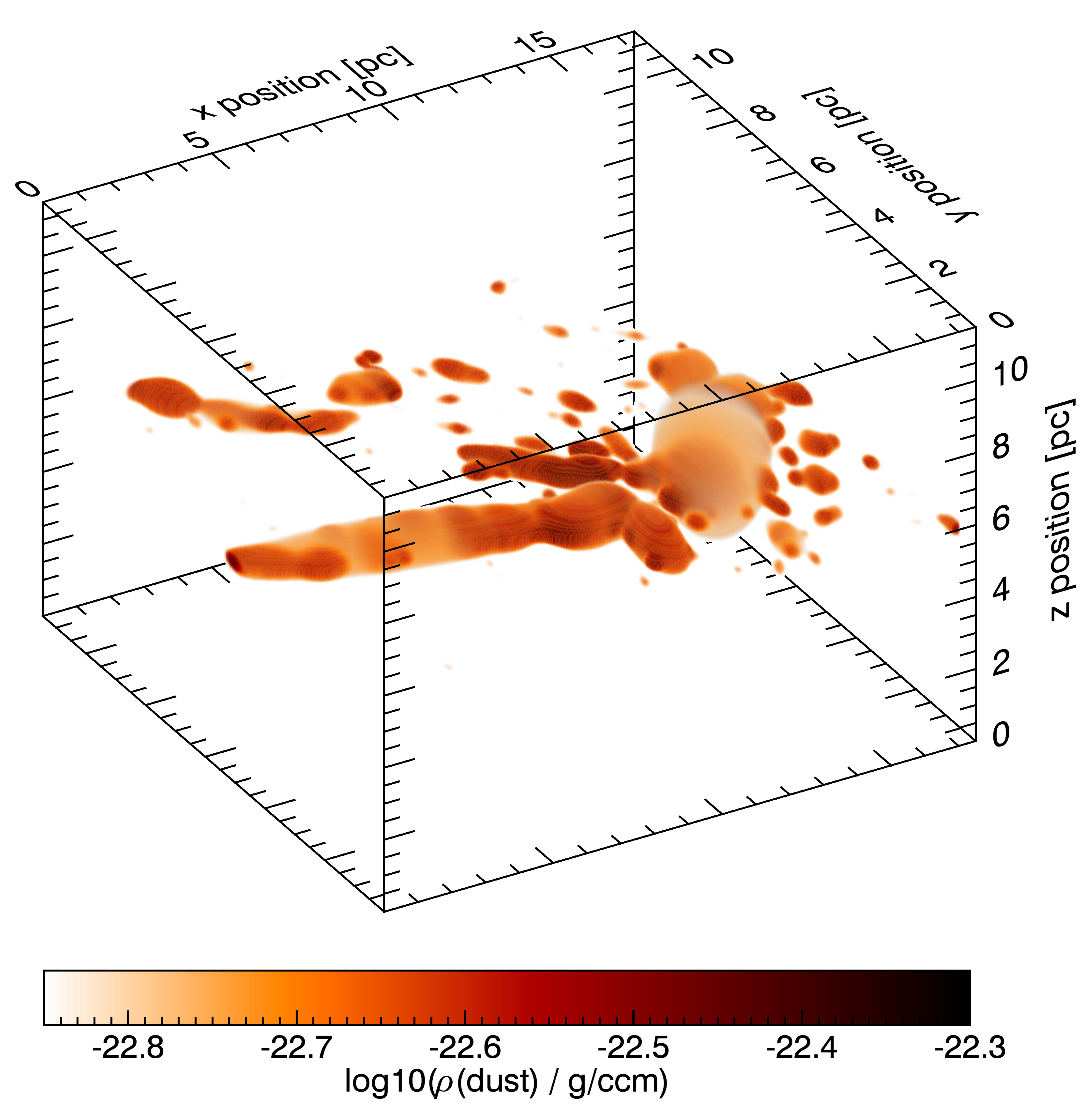}
			\subcaption{$\rho$ Ophiuchi cloud. }
			\label{pic_appli_rhooph}
		\end{subfigure}
		\hfill
		\begin{subfigure}[c]{0.485\textwidth}
			\includegraphics[width=\textwidth]{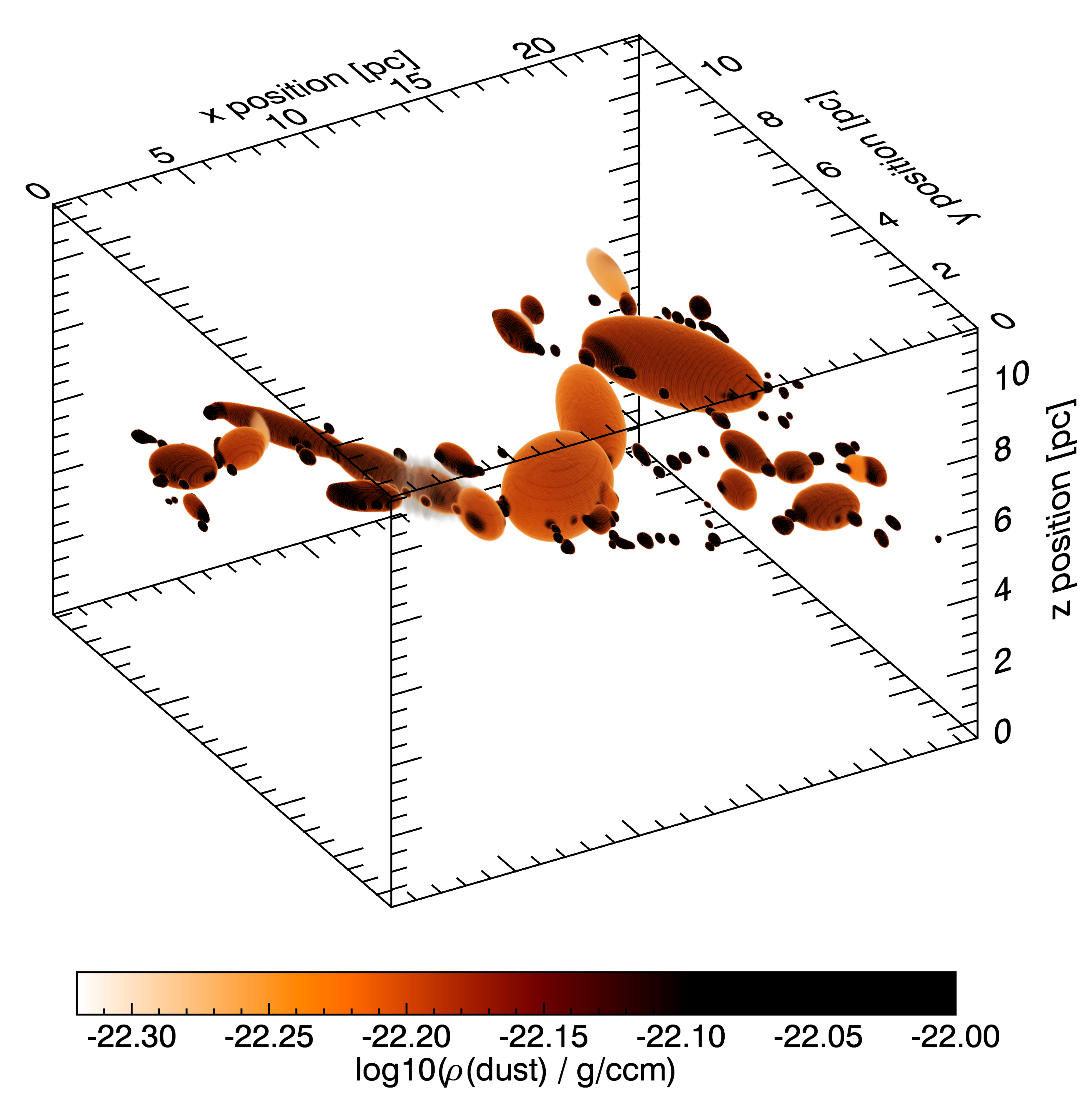}
			\subcaption{G11.11 filament, the \textit{Snake}. }
			\label{pic_appli_g11}
		\end{subfigure}
		\caption{3D images of the dust density distribution of the $\rho$ Ophiuchi \citep[based on][]{Kainulainen2009} and G11.11 models \citep[based on][]{Kainulainen2013b}. 
			The models consist of a number of small, dense clumps that are surrounded by diffuse envelopes. 
			Note that we only plot the most dense parts of the models and not the more diffuse envelopes since those would obscure the central clumps and cores.
			The real range of total column densities and dust temperatures are summarised in Table \ref{tab_models_derived}.
		}
		\label{pic_appli}
	\end{figure*}

	Next, we study the sight line effects on models that are more complex in both their density and temperature structures and represent real filaments better than the cylinders in Sect.\ \ref{analytic} do.
	We use MC to derive the dust temperature and emission in the FIR and sub-mm self-consistently (contrary to the isothermal models we studied in Sect.\ \ref{analytic}).
	We discuss two main questions:
	Firstly, how would we observe filamentary structures at different viewing angles?
	Secondly, are there observational criteria, such as the dust temperature and total column density, which enable us to reconstruct elongated structures along the LoS by dust observations alone?

	We use two 3D models of molecular clouds that estimate the volume density distributions by fitting a superposition of spheroids on observed column density maps.
	Fig.\ \ref{pic_appli} shows 3D images of our models, one based on the $\rho$ Ophiuchi cloud \linebreak \citep[Fig.\ \ref{pic_appli_rhooph},][]{Kainulainen2009,Kainulainen2014} and the other on the G11.11 \textit{Snake} filament \citep[Fig.\ \ref{pic_appli_g11},][]{Kainulainen2013b,Kainulainen2014}.
	In both models, the x-y-plane is the plane of the sky and the z-axis represents the LoS.
	Table \ref{tab_models_details} lists the basic parameters we used for our simulations and Table \ref{tab_models_derived} provides a summary of the volume and column density, dust temperature, and mean flux density statistics of both models.

	\footnotetext[1]{Note that we give values for the gas density here. The code, however, requires the dust density as input that we derive by assuming a gas-to-dust mass ratio of $R$ = 100.}

	We derive dust temperatures, assuming that the dust is only heated by an external, isotropic heating field.
	We use the model of interstellar radiation field (ISRF) by \citet{Mathis1983}.
	Their ISRF consists of a combination of three diluted black-body spectra, with effective temperatures of 7500 K, \linebreak 4000 K, and 3000 K, and an additional UV excess \citep{Mathis1983}. 
	We produce images by ray-tracing \citep[see ][]{Heymann2012} at 881 wavelengths within a range from 0.43 \AA{} to 1.2 cm.
	To reduce the MC noise we use 5 $\times$ 10$^4$ and 10$^6$ photon packages per frequency grid point for the $\rho$ Ophiuchi and \textit{Snake} model, respectively. 
	Previous tests have demonstrated that this is sufficient to sample the temperature within all cells continuously, which is reflected by a noise-less spectrum.

	Using only the ISRF as the heating source is a simplification.
	The $\rho$ Ophiuchi cloud, for example, is a site of active star formation \citep{Wilking2008} that heats the cloud internally, which leads to significant temperature gradients.
	Recognising this caveat, we focus on a simple external heating field only.

	As a dust model we use a mixture of amorphous carbon (aC) and silicates (Si) dust grains.
	The grain sizes range between \linebreak 16 to 128 nm and 32 to 256 nm for aC and Si grains, respectively, and follow the size distribution with number density $\propto \, a^{-3.5}$ d$a$ \citep{MRN1977}.
	The cross-sections and opacities are based on the dust models of \citet{Siebenmorgen2014}.
	We choose spherical grains and abundances relative to H$_2$ of \linebreak $\chi_{\rm aC}$ = 2.5 $\times$ 10$^{-3}$ and $\chi_{\rm Si}$ = 4.8 $\times$ 10$^{-3}$, respectively.
	This ratio is consistent with typical values of dust in the interstellar medium in the solar neighbourhood \citep{Dwek2005}.

	We are aware that our dust grain model is a simplification, as well.
	Studies by, for example, \citet{Ossenkopf1994}, \citet{Stepnik2003}, \citet{Steinacker2015}, \citet{Lefevre2016} indicate that dust grains in dense parts of the ISM grow to larger aggregates.
	In the future, we will extend the parameter space to study the influence of other grain models.
	For now, we keep our simplified set-up and focus on analysing filaments at different viewing angles.

	We consider 40 different directions by varying the inclination, $\theta$, between 0$^\circ$ and 315$^\circ$ and position angle, $\varphi$, between 0$^\circ$ and 180$^\circ$, both in steps of 45$^\circ$.
	We focus our analysis on the FIR and sub-mm part of the spectrum since this part is dominated by the thermal emission of dust.
	To mimic Herschel observations at the \textit{Herschel} SPIRE bands (250, 350, and 500 $\mu$m) we smooth the flux density maps to the resolution of the SPIRE 500 $\mu$m data using Gaussian point-spread functions.
	For each pixel of the synthetic intensity maps, we fit a modified black body to the flux values at \linebreak $\lambda$ = 250, 350 and 500 $\mu$m, following the descriptions in \citet[][see Eq.\ (\ref{def_mbnu})]{Koenyves2010}.
	We assume a mass absorption coefficient $\kappa_{\nu}$ = 0.1 cm$^2$ g$^{-1}$ $\left( \frac{\nu}{\mbox{1000 GHz}} \right)^\beta$.
	For the dust emissivity index, $\beta$, we use 1.75 \citep{Koenyves2010,Wang2015}.
	The results we discuss in the following sections are based on this value.
	We have performed the same analysis with $\beta$ = 2.0 and recover similar results within the typical observational uncertainties.

	We use the \texttt{idl} routine \texttt{mpfitfun} \citep{Markwardt2009} to perform the fittings, with the effective dust temperature and total column density as parameters. 
	The resulting maps are shown in Appendix \ref{app_images}.
	Fig.\ \ref{pic_rhooph_example} presents results based on the $\rho$ Ophiuchi model, including flux density, effective column density, and effective dust temperature maps at ($\theta$,$\varphi$) = (0$^\circ$,0$^\circ$) and (45$^\circ$,45$^\circ$).

	We note that $T_{\rm d}^{\rm eff}$ and $N_{\rm tot}^{\rm eff}$ are not direct results of the RT simulations, but derived by SED fitting of three data points and assuming a power-law opacity relation ($\kappa \, \propto \, \nu^\beta$).
	This method introduces some degree of degeneracy.
	For this reason both quantities are less accurate than the values that result from the RT procedure.
	We find a correlation between the real column density, which we computed directly from the input volume density distribution, and the effective column density derived by SED fitting which is lost by some scatter owing to the inaccuracies of fitting an entire SED to three points with only one single temperature.

	However, $T_{\rm d}^{\rm eff}$ and $N_{\rm tot}^{\rm eff}$ are equivalent to dust temperatures and column densities that are derived from observed fluxes.

	\subsection{$\rho$ Ophiuchi model}\label{apply_rhooph}

	\begin{mysidewaysfigure}
		\includegraphics[width=\textwidth]{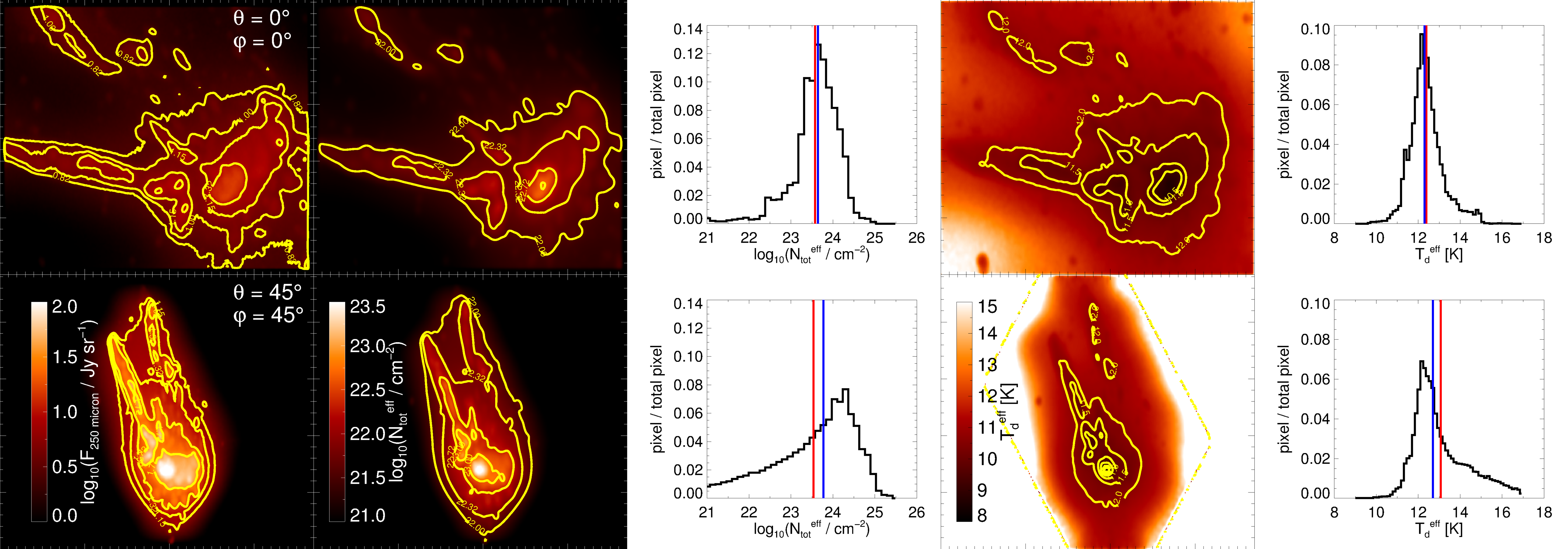}
		\caption{Results based on the $\rho$ Ophiuchi cloud model at ($\theta$,$\varphi$) = (0$^\circ$,0$^\circ$) (\textit{top}) and ($\theta$,$\varphi$) = (45$^\circ$,45$^\circ$) (\textit{bottom}).
			From \textit{left} to \textit{right} the figures show the maps of synthetic intensity at 250 $\mu$m, maps of effective total column density with corresponding PDFs, and the maps of effective dust temperature with corresponding PDF.
			The contours in the maps are at $\log_{\rm 10}\left(F_{\rm 250 \mu m} \, / \, {\rm Jy \, beam}^{-1} \right) = $ \{0.82,1.0,1.15,1.32,1.5\}, $\log_{\rm 10}\left(N_{\rm tot}^{\rm eff} \, / \, {\rm cm}^{-2} \right) = $ \{22.0,22.32,22.72,23.0\}, and $T_{\rm d}^{\rm eff} = $ \{10.5,11.0,11.5,12.0\} K.
			The red lines in the PDFs indicate the arithmetic mean values of the distributions, the blue lines the median values.
		}
		\label{pic_rhooph_example}
	\end{mysidewaysfigure}

	\begin{figure*}[htb]
		\centering
		\includegraphics[height=0.37\textheight]{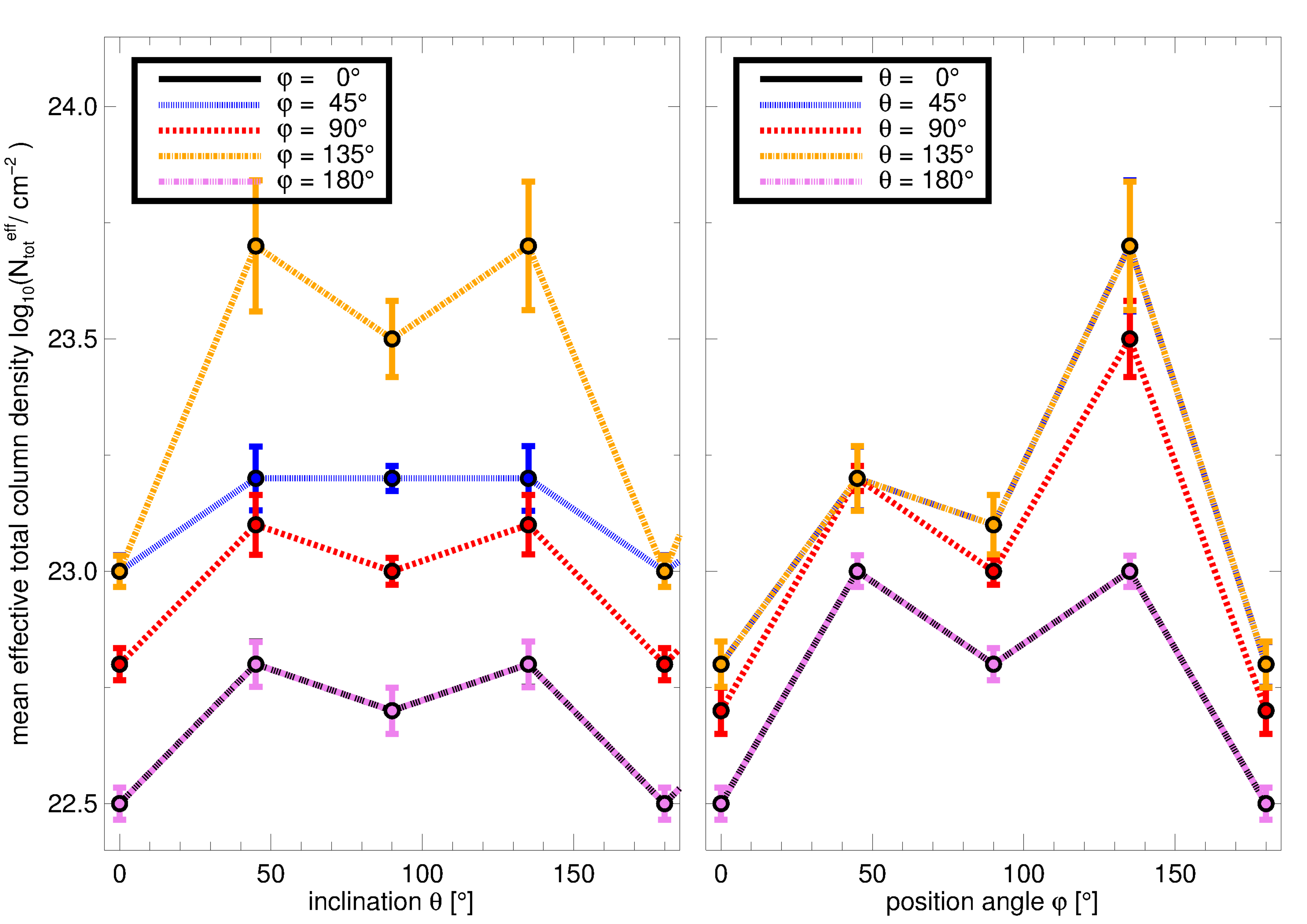}
		\caption{$\rho$ Ophiuchi cloud model.
			The plots show the mean values of the effective total column density PDFs, $\xbar{N}_{\rm tot}^{\rm eff}$, as a function of the inclination, $\theta$, and position angle, $\varphi$.
			The error bars illustrate 1$\sigma$ errors.
			Note that we only show the values for inclinations ${\theta \, \leq\rm{ 180}^\circ}$ because the values for $\theta$ and $\theta$+180$^\circ$ are identical.
			We see that the mean dust column density strongly depends on the viewing angle.
			The maximal difference compared to ($\theta$,$\varphi$) = (0$^\circ$,0$^\circ$) is on the order of 1.7 dex.
			}
		\label{pic_rhooph_histostat_nmap_mean}

		\includegraphics[height=0.37\textheight]{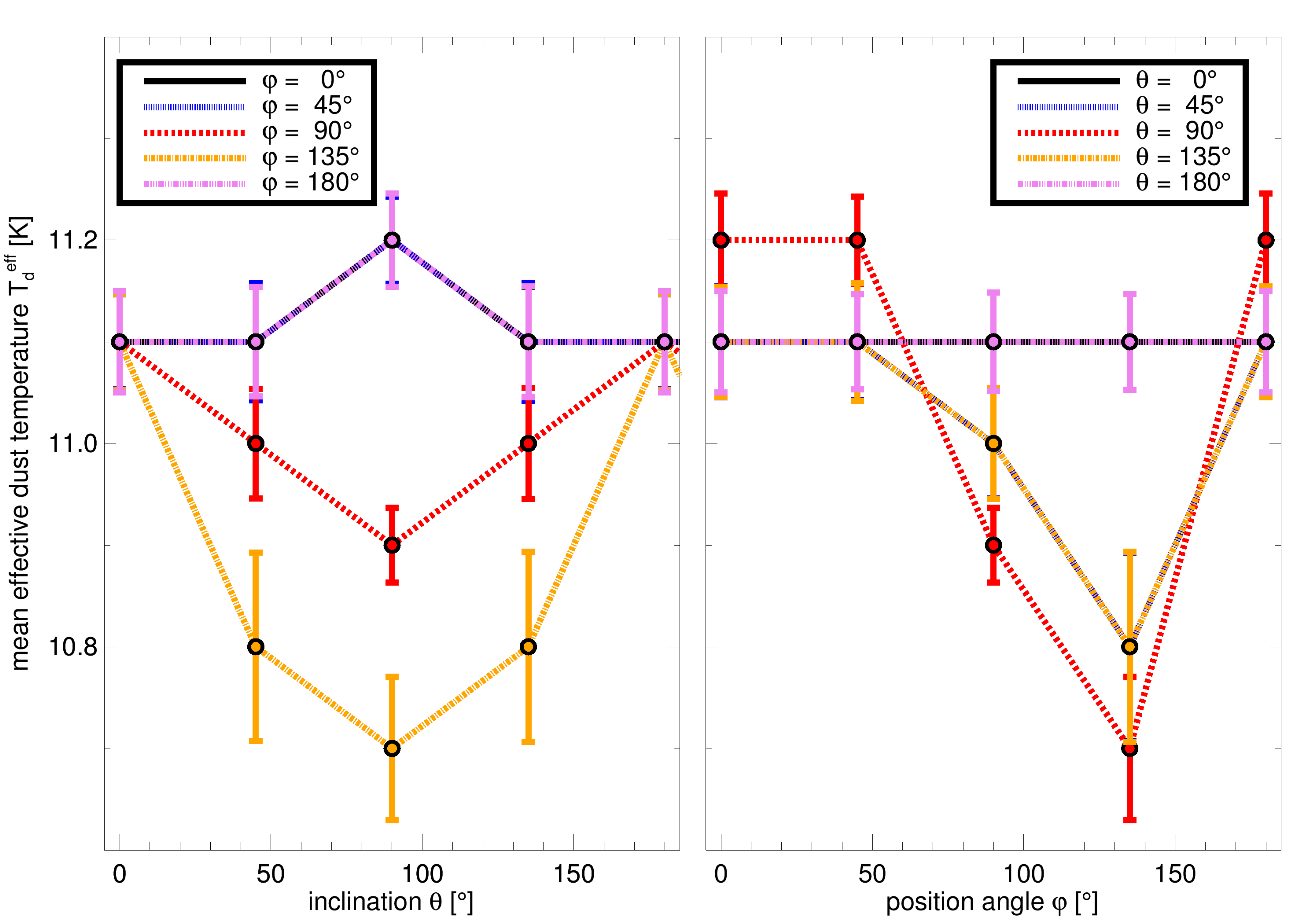}
		\caption{$\rho$ Ophiuchi cloud model.
			Like Fig.\ \ref{pic_rhooph_histostat_nmap_mean} relating to the mean values of effective dust temperature PDFs.
			$\xbar{T}_{\rm d}^{\rm eff}$ does not depend strongly on the viewing angle.
			}
		\label{pic_rhooph_histostat_tmap_mean}
	\end{figure*}

	Figures \ref{pic_rhooph_tmap} and \ref{pic_rhooph_nmap} show maps of effective dust temperature and total column density respectively, based on the $\rho$ Ophiuchi model.
	We find similar results along a given LoS \linebreak (front $\rightarrow$ back compared to back $\rightarrow$ front), just mirror-inverted (e.\ g., maps at ($\theta$,$\varphi$) = (0$^\circ$,0$^\circ$) and (180$^\circ$,0$^\circ$)).
	As long as the dust is optically thin we see the same profiles both ways.
	For molecular lines, however, the sight line will make a significant difference \citep[e.g.,][]{Chira2014}.

	Looking at our model along different LoSs, we see significant changes in the morphology. 
	The more we incline the object the clumpier and more elliptic the morphology becomes.
	This is expected based on the structure of the input model.

	We observe that the effective total column density of the central region becomes higher while its gradient becomes steeper the more we incline the model.
	The central effective dust temperatures, however, do not change significantly, whereas the profile becomes flatter.
	Fig.\ \ref{pic_rhooph_example} shows a detailed example in which we plot maps of synthetic 250 $\mu$m flux density, $F_{\rm 250 \mu m}$, the effective total column density, $N_{\rm tot}^{\rm eff}$, and effective dust temperature, $T_{\rm d}^{\rm eff}$, as seen at ($\theta$,$\varphi$) = (0$^\circ$,0$^\circ$) and ($\theta$,$\varphi$) = (45$^\circ$,45$^\circ$), as well as the respective PDFs.
	We note that since the dust is optically thin in the FIR the flux and column densities behave in the same way.
	This is why we focus our discussion on $N_{\rm tot}^{\rm eff}$.

	The effective column density is concentrated towards the longest axis of the model at ($\theta$,$\varphi$) = (45$^\circ$,45$^\circ$) compared to \linebreak ($\theta$,$\varphi$) = (0$^\circ$,0$^\circ$), leading to higher intensities at similar effective dust temperatures.
	If one compares the regions enclosed by the contours at both viewing angles one can identify the same structures at ($\theta$,$\varphi$) = (45$^\circ$,45$^\circ$) as in ($\theta$,$\varphi$) = (0$^\circ$,0$^\circ$) (considering the influence of to rotation), but at higher values in the case of the intensity and effective column density.
	In the case of effective dust temperature, the respective regions are indicated by the contours of the same values. 

	At ($\theta$,$\varphi$) = (0$^\circ$,0$^\circ$), the PDFs for both the effective total column density and dust temperature are log-normal with a peak at $\xbar{N}_{\rm tot}^{\rm eff}$ = 4.4 $\times$ 10$^{21}$ cm$^{-2}$ and $\xbar{T}_{\rm d}^{\rm eff}$ = 12.2 K.
	The PDFs of \linebreak ($\theta$,$\varphi$) = (45$^\circ$,45$^\circ$), however, have the shape of a log-normal distribution with a power-law tail towards lower column densities and higher dust temperatures, peaking at $\xbar{N}_{\rm tot}^{\rm eff}$ = 2.0 $\times$ 10$^{22}$ cm$^{-2}$ and $\xbar{T}_{\rm d}^{\rm eff}$ = 12.2 K.

	The main point that can be deduced from this result is that observing filaments along different sight lined do not fake typical signatures of star formation in column density PDFs \linebreak (cf.\ \citealt{Schneider2013,Stutz2015}).
	The tails towards lower column densities in the PDFs are normally within the noise range and cut during the data reduction process.
	We discuss this in Sect.\ \ref{apply_g11} in more detail.

	In Figs.\ \ref{pic_rhooph_histostat_nmap_mean} and \ref{pic_rhooph_histostat_tmap_mean} we plot the values of $\xbar{N}_{\rm tot}^{\rm eff}$ and $\xbar{T}_{\rm d}^{\rm eff}$ from fitting Rayleigh distributions to the PDFs (see Sect.\ \ref{analytic}) as a function of viewing angle.
	We find huge variations in $\xbar{N}_{\rm tot}^{\rm eff}$ compared to the non-inclined sight line ($\theta$,$\varphi$) = (0$^\circ$,0$^\circ$).
	The maximal $\xbar{N}_{\rm tot}^{\rm eff}$ is detected at ($\theta$,$\varphi$) = (45$^\circ$,135$^\circ$) and (135$^\circ$,135$^\circ$) at a factor of 50 higher than the ($\theta$,$\varphi$) = (0$^\circ$,0$^\circ$) case.
	These average column densities are clearly above the average range of mean total column density in the literature \citep[5 $\times$ 10$^{20}$ -- 3.16 $\times$ 10$^{22}$ cm$^{-2}$][]{Koenyves2010,Juvela2012b,Palmeirim2013,Roy2014}.

	In contrast, the variations of $\xbar{T}_{\rm d}^{\rm eff}$ are below 0.5 K.
	These differences are on the same order as typical noise levels and would not be significant enough to make any profound statement about the viewing angle.
	For a fixed $\theta$ = 0$^\circ$, the $\xbar{T}_{\rm d}^{\rm eff}$ remains constant.
	All values for $\xbar{T}_{\rm d}^{\rm eff}$ here are within a range that agrees with observational findings of pre-stellar cores and filaments \citep[10 -- 15 K, e.\ g.][]{Juvela2012b}.

	Comparing the results of the $\rho$ Ophiuchi model with those in Sect. \ref{analytic}, the closest model with respect to the density profiles would be cylsph.
	We argue that the differences in behaviour result from the fact that the $\rho$ Ophiuchi model is dominated by its irregular, clumpy structure which makes it more sensitive to changes in the viewing angles than a smooth, regular structure.
	Furthermore, the total model of $\rho$ Ophiuchi is not isothermal, unlike the individual spheroids.
	Since the individual spheroids are approximately homogeneous, the $\rho$ Ophiuchi model can be described as a superposition of homcyl-like elements which, however, has not shown significant variations along different sight lines as we observe them for the $\rho$ Ophiuchi model.

	Thus, more complicated structures are more significantly affected by changes in orientation and geometry.
	However, the column density alone is not enough to trace back the inclination of an observed filament.
	Only a significant increase in column density at a constant average dust temperature can give hints on how the matter is distributed along a given LoS.

	Just inclining the model does not change the thermal processes (heating, cooling) within the object. 
	If the density within the object increases (for example, by collapsing material) the dust is cooled more efficiently and $T_{\rm d}^{\rm eff}$ decreases before the first protostar is formed.
	Thus, if $N_{\rm tot}^{\rm eff}$ and $T_{\rm d}^{\rm eff}$ do not behave anti-proportionally, this indicates that the object is elongated along the LoS.
	In this case, it is necessary to re-construct the density distribution along the LoS by line observations.

	\subsection{G11.11 \textit{Snake} model}\label{apply_g11}

	\begin{figure*}[htb]
		\centering
		\includegraphics[height=0.39\textheight]{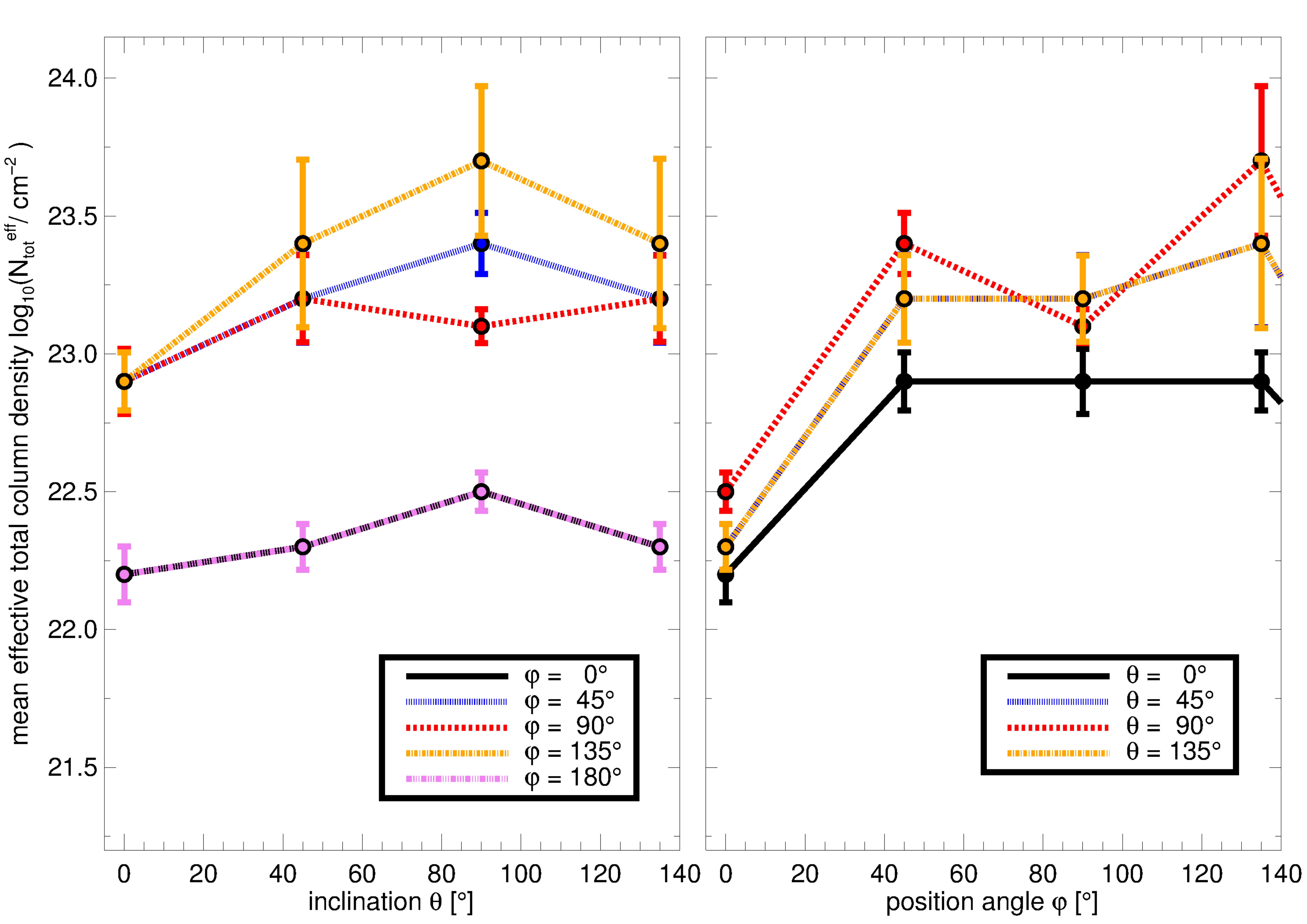}
		\caption{Like Fig.\ \ref{pic_rhooph_histostat_nmap_mean} but for the G11.11 \textit{Snake} model.
			We see a strong dependence of the $\xbar{N}_{\rm tot}^{\rm eff}$ on the viewing angle.
			The differences compared to ($\theta$,$\varphi$) = (0$^\circ$,0$^\circ$) are on the order of $\sim$ 1.2 -- 2.5 dex.
		}
		\label{pic_g11_histostat_nmap_mean}

		\includegraphics[height=0.39\textheight]{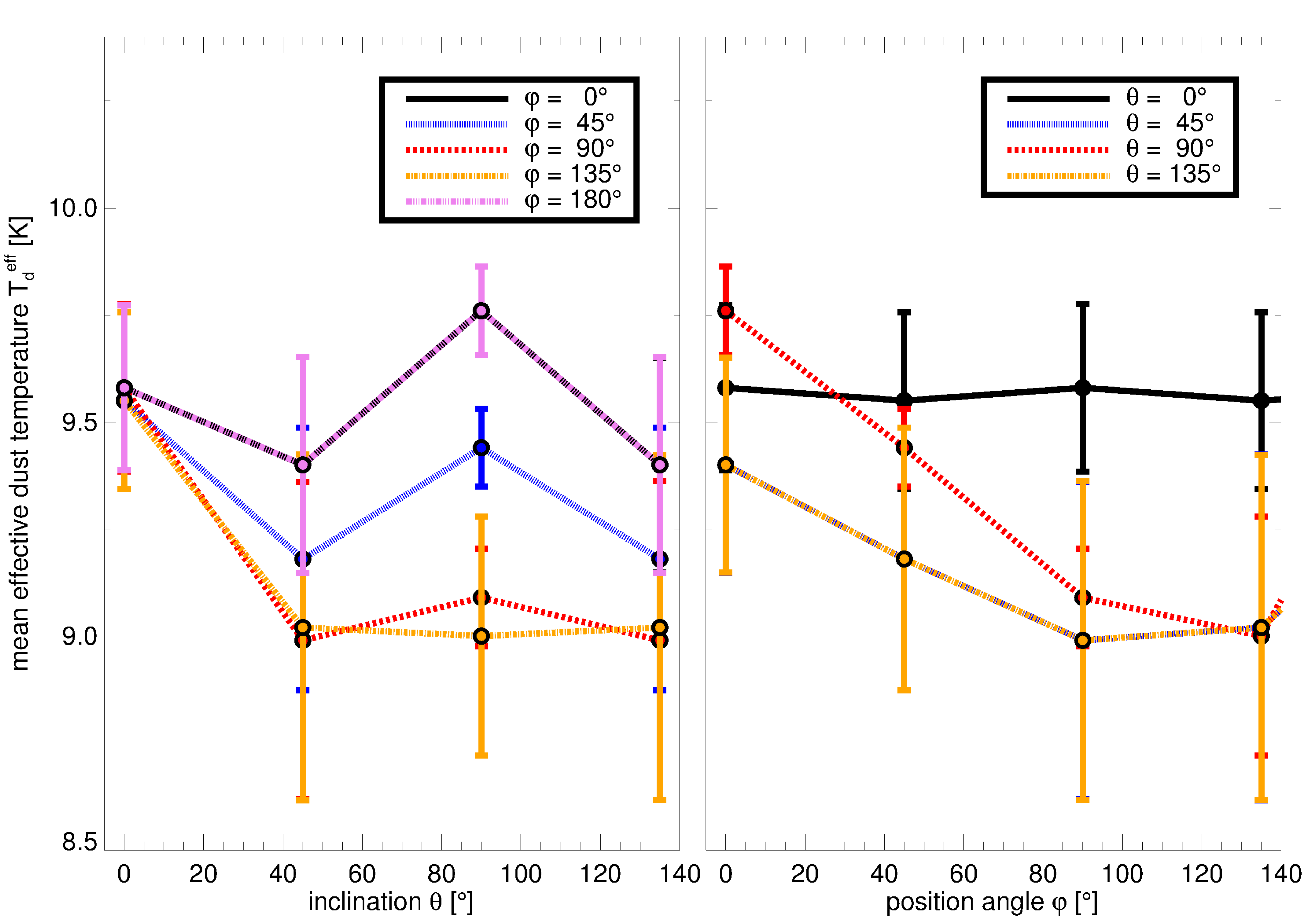}
		\caption{Like Fig.\ \ref{pic_rhooph_histostat_tmap_mean} but for the G11.11 \textit{Snake} model.
		}
		\label{pic_g11_histostat_tmap_mean}
	\end{figure*}

	\begin{figure*}
		\centering
		\includegraphics[width=\textwidth]{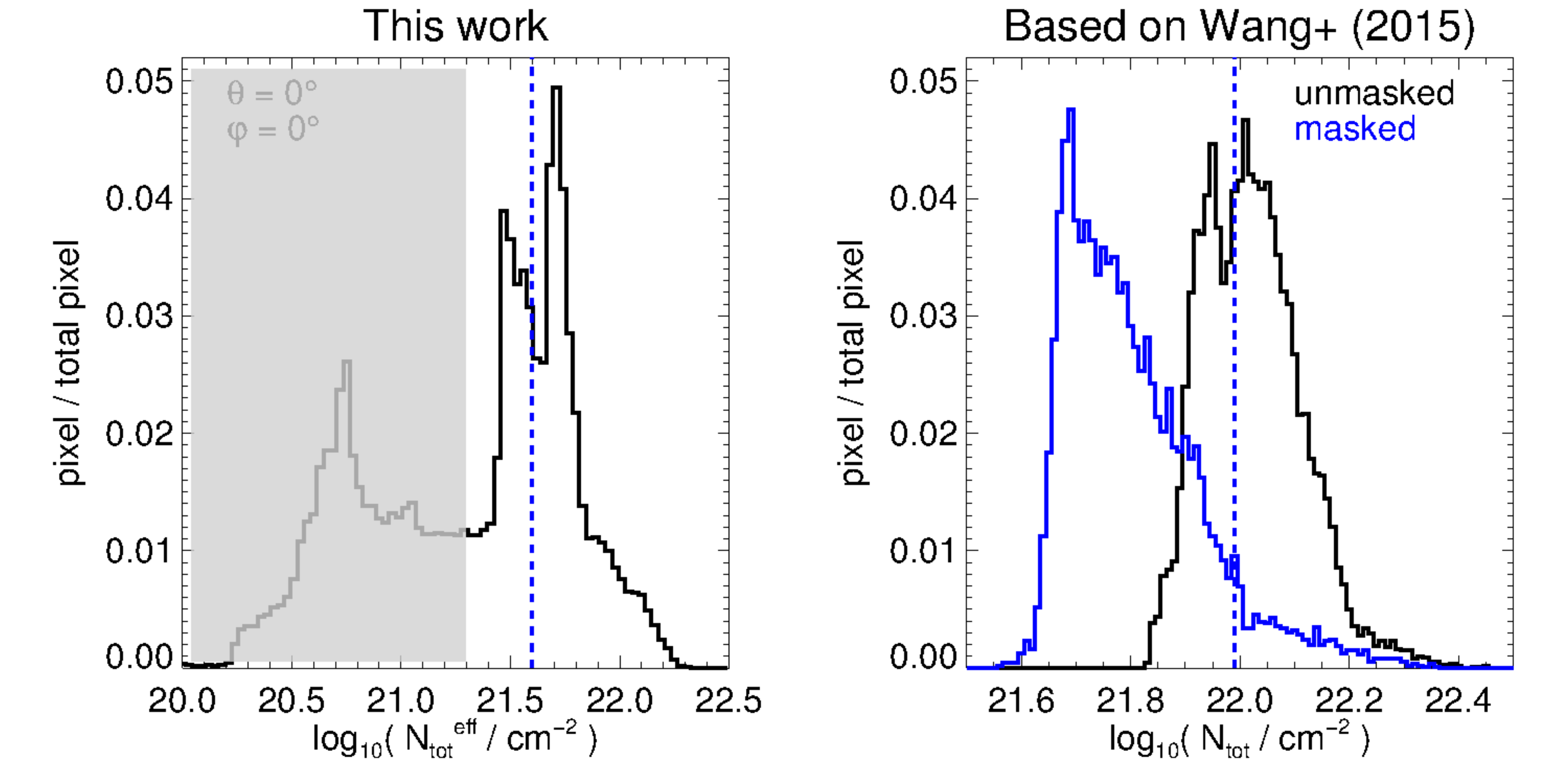}
		\caption{Comparison of (effective) total column density PDFs of \textit{Snake} based on data of this work (\textit{left}) and \citet[\textit{right}]{Wang2015}.
			The black lines show the original data of the simulation and observation, respectively. 
			The grey-shaded area on the left plot indicates the range that would be below the detection limit of \textit{Herschel}.
			The blue-dotted line in both images indicates the cut limit which is used for noise reduction.
			The solid blue line in the right plot shows the total column density PDF based on the masked intensity maps.
		}
		\label{pic_g11_comp_wang}
	\end{figure*}

	\begin{figure*}[htb]
		\centering
		\includegraphics[height=0.39\textheight]{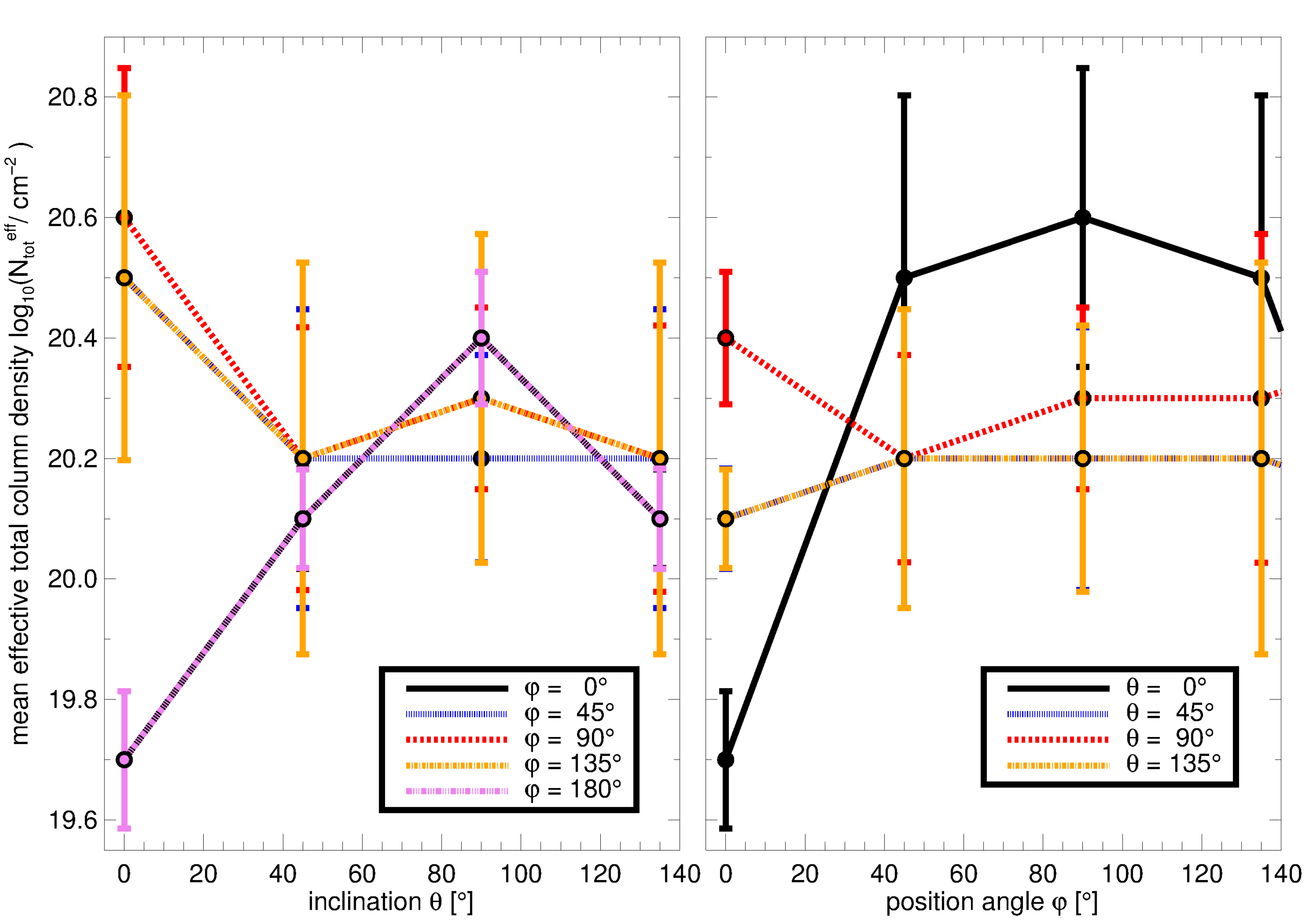}
		\caption{Like Fig.\ \ref{pic_g11_histostat_nmap_mean} but based on the masked maps of the G11.11 \textit{Snake} model.
			We see that the behaviour here is, in general, the same as before, just the values and differences are lower due to the intensity cut.
		}
		\label{pic_g11_intcut_histostat_nmap_mean}

		\includegraphics[height=0.39\textheight]{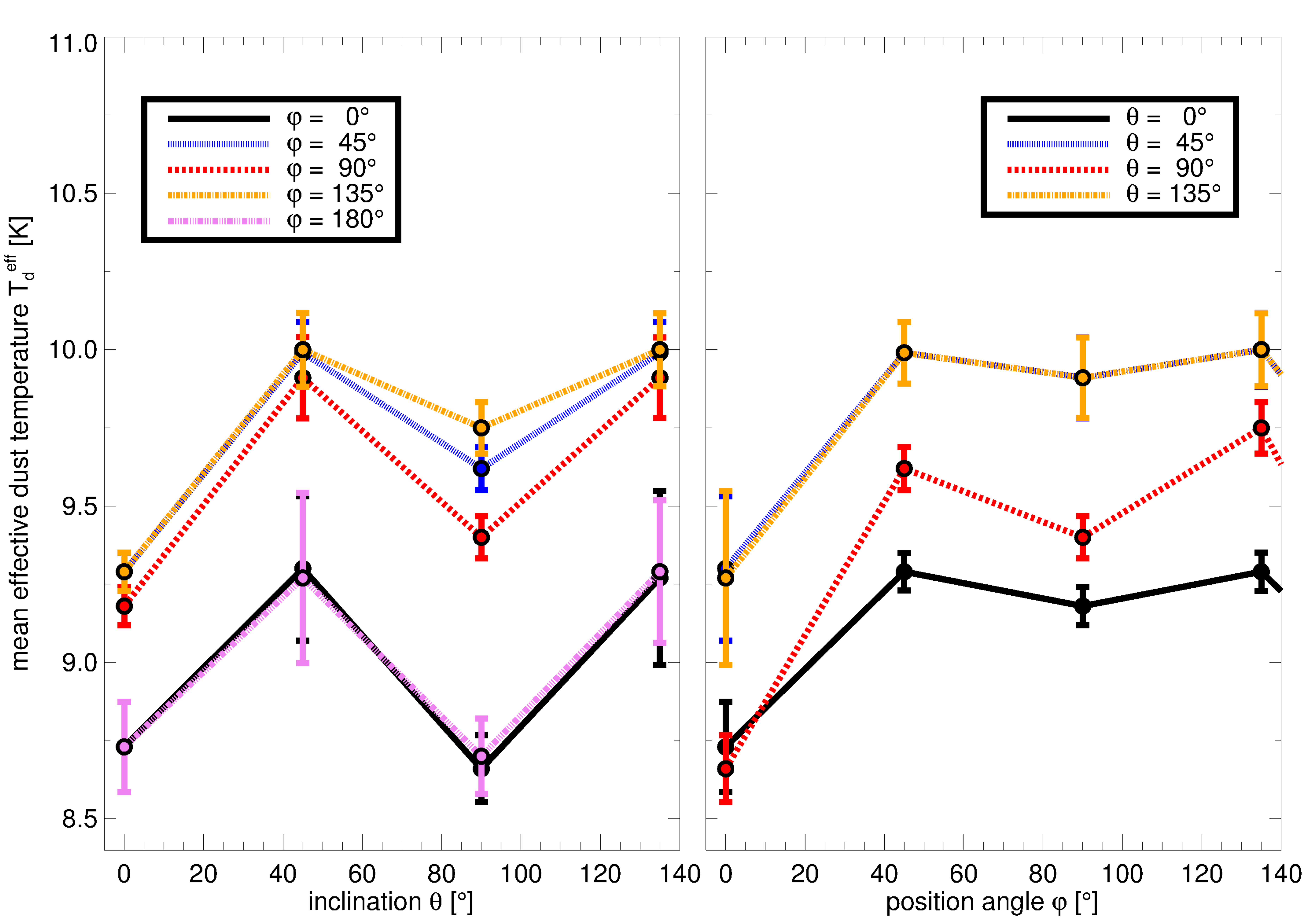}
		\caption{Like Fig.\ \ref{pic_g11_histostat_tmap_mean} but based on the masked maps of G11.11 \textit{Snake} model.
			We see that the $\xbar{T}_{\rm d}^{\rm eff}$ is approximately constant for all viewing angles, but variations between different sight lines are larger.
		}
		\label{pic_g11_intcut_histostat_tmap_mean}
	\end{figure*}

	We repeat the analysis for the \textit{Snake} model and show the maps of effective dust temperature and total column density in Figs.\ \ref{pic_g11_tmap} and \ref{pic_g11_nmap}, respectively.
	We note that we only print the maps within (0$^\circ \,\leq \, \theta \, \leq$ 135$^\circ$, 0$^\circ \, \leq \varphi \, \leq$ 180$^\circ$) owing to the symmetries we have already discussed in Sect.\ \ref{apply_rhooph}.

	We observe that rotations do not skew the projected morphology of \textit{Snake} as much as it has done for $\rho$ Ophiuchi.
	At all viewing angles (except $\theta$ = 90$^\circ$), we can see the sinusoidal structure which is typical for the \textit{Snake}, although its amplitude slightly changes.
	The distribution of effective dust temperature, however, remains approximately constant, as has been the case for $\rho$ Ophiuchi.
	The values of the average effective dust temperature $\xbar{T}_{\rm d}^{\rm eff}$ (see Fig.\ \ref{pic_g11_histostat_tmap_mean}) vary within 0.8 K, which is insignificant in terms of observational uncertainties \citep[e.\ g.,][]{Juvela2012b}.

	As with the $\rho$ Ophiuchi model, we see significant variations in effective total column density with viewing angle.
	As plotted in Fig.\ \ref{pic_g11_histostat_nmap_mean}, the values of $\xbar{N}_{\rm tot}^{\rm eff}$ at ($\theta$,$\varphi$) = (0$^\circ$,0$^\circ$) and \linebreak ($\theta$,$\varphi$) = (0$^\circ$,45$^\circ$) change by almost two orders of magnitude.
	In general, we observe an increase in $\xbar{N}_{\rm tot}^{\rm eff}$ by factors of between \linebreak 16 and 320 compared to ($\theta$,$\varphi$) = (0$^\circ$,0$^\circ$).
	The steepest rise is found when rotating the model between ($\theta$,$\varphi$) = (0$^\circ$,0$^\circ$) to positions with $\theta$ = 45$^\circ$ or $\varphi$ = 45$^\circ$.
	For steeper angles, the column densities do not change as significantly.

	Thus, the distributions of effective total column density and dust temperature of the \textit{Snake} model behave similarly to our $\rho$ Ophiuchi model.
	The elongated structure of the model amplifies the variations of the small components in two ways;
	firstly, the $N_{\rm tot}^{\rm eff}$ rises higher along the LoS, which is parallel to its long axis than along other sight lines;
	secondly, the radial temperature gradient is larger since the filament is too thin and not dense enough to be isothermal.

	Compared to the results in Sect.\ \ref{apply_rhooph}, we note that the errors of the mean values of the distributions are larger than before.
	We can explain that by having a closer look at the PDFs, for which Fig.\ \ref{pic_g11_comp_wang} shows an example.
	In the left panel we plot the effective total column density PDF (black line) of our \textit{Snake} model as observed at ($\theta$,$\varphi$) = (0$^\circ$,0$^\circ$).
	We see that the PDF splits into three peaks in total; two at higher column densities \linebreak ($\sim$ 3.2 -- 5 $\times$ 10$^{21}$ cm$^{-2}$), which are relatively close to each other and a third peak at low column densities ($\sim$ 5 $\times$ 10$^{20}$ cm$^{-2}$).
	A similar profile is seen with the effective dust temperature PDF. 
	Since we only fit a single Rayleigh distribution to the PDF, this is a major source of the uncertainties. 

	Of course, we could reduce the errors by fitting a super-position of Rayleigh distributions for each component.
	However, in terms of physical interpretation that would mean that we assume that the model consists of at least three independent components.
	Furthermore, finding these peaks has already been unexpected since there is no hint in the literature that this structure has been observed for the real \textit{Snake} cloud.

	We conclude that this is due to observational limitations and reduction processes (Ke Wang, private communication).
	On the right side of Fig.\ \ref{pic_g11_comp_wang}, we plot the total column density PDF derived by \citet{Wang2015}.
	The black solid line shows the column density PDF derived from the unmasked (calibrated, but not noise-reduced) intensity map, and with the blue solid line the PDF for the masked map.
	The blue dashed line indicates the level at which the originally observed intensity maps have been cut in the masking process.

	For the \citeauthor{Wang2015} study, fluxes corresponding to total column densities below 2.5 $\times$ 10$^{21}$ cm$^{-2}$ are below the sensitivity limit of \textit{Herschel} and would not be observed.
	We indicate this part with a grey box in our PDF (Fig.\ \ref{pic_g11_comp_wang}, left plot).
	This means that we could detect the two peaks at higher column densities with observations, but not the third one at low column densities.
	This is consistent with the fitting routine of \citet{Kainulainen2014}, which weights the highest column density regimes more than the low column density parts.
	Thus, the lower peak belongs to the parts of the observations that lie within the noise and is cut off during the noise correction process.
	Applying all these factors, we end up with a single-peak PDF which agrees with the observed findings.

	We note that the peak positions in the PDFs of \citeauthor{Wang2015} and this work are not the same.
	First, our model filaments were derived from extinction maps in the NIR, whereas \citet{Wang2015} used emission maps in the sub-mm.
	Even though these techniques result in column densities that on average are in agreement \citep[e.g.,][]{Kainulainen2013b}, regional differences may exist.
	On the one hand, extinction maps are limited at high column densities whereas emission maps are most sensitive there.
	On the other hand, emission maps are more temperature-dependent than extinction maps.

	Second, our models are only heated externally.
	Both \citet{Henning2010} and \citet{Wang2015} show that this does not match reality.
	The \textit{Snake} filament contains cores that are associated with protostars and ongoing star formation processes.
	While the results of cylsph in Sect.\ \ref{analytic} suggest that missing individual cores does not affect the column density statistics of the filament, a larger population of cores would contribute to the incoming radiation field if they contain embedded sources and thus change the outputs of the RT code.
	Moreover, we do not consider the temperature gradient in cylsph, which would be expected for the \textit{Snake} filament.
	The influence of different external heating fields and/or internal heating sources needs to be investigated in more detail in future studies.

	We test the significance of noise by cutting the noise level from our flux maps and then repeat the analysis of the resulting PDFs.
	The results for the $\xbar{N}_{\rm tot}^{\rm eff}$ and $\xbar{T}_{\rm d}^{\rm eff}$ are plotted in \linebreak Figs.\ \ref{pic_g11_intcut_histostat_nmap_mean} and \ref{pic_g11_intcut_histostat_tmap_mean}, respectively.
	We find that the effective total column density are lower than before, by about one order of magnitude, whereas the effective dust temperatures keep relatively constant although the variations become larger.

\section{Conclusions}\label{conclusions}

	In this paper, we have investigated the dependence of effective total column density and dust temperature of filaments on the viewing angle.

	As a first step, we have analysed the behaviour of flux density profiles of simple cylinders as a function of viewing angle and dust temperature.
	We have found that the mean flux density rises with increasing dust temperature, but it is not significantly influenced by changing the viewing angle.

	Next, we have analysed 3D models of the $\rho$ Ophiuchi cloud and the G11.11 \textit{Snake} filament and performed radiative transfer calculations to compute images for 40 different viewing angles.
	We have focused our analysis on the frequencies that are within the \textit{Herschel} SPIRE bands and derived the effective dust temperature, $T_{\rm d}^{\rm eff}$, and total column density, $N_{\rm tot}^{\rm eff}$, for each pixel by SED fitting.

	We have found that the $\xbar{T}_{\rm d}^{\rm eff}$ is approximately constant, whereas $\xbar{N}_{\rm tot}^{\rm eff}$ changes significantly, especially when we rotate the models into the direction where their long axis is parallel to the LoS.
	Since the dust emission is optically thin, the column densities strongly depend on the viewing angle.
	The dust temperature is determined by the local heating and cooling processes, which is unchanged when we rotate the models.

	We have briefly investigated how sensitivity limits and noise corrections influence our results.
	We have found that common data reduction processes reduce the level of variation in column density.
	The variations in effective dust temperature have increased, but have been still insignificant enough in the observational context.

	We conclude that there is no quantity in our analysis related to dust emission that tracks the inclination of a filament uniquely.
	A notably high column density at a given dust temperature can indicate that an observed object is elongated along the LoS direction.
	For true inclinations and confirming masses, line observations are required.
	However, with all the data obtained by dust surveys and our findings, it is possible to identify candidates of filaments which may be elongated along the LoS.
	This is important to learn more about the distribution and orientation of filaments in the Galactic plane, which improves our understanding of the role of filaments within the star formation process.

	\begin{acknowledgements}
		The authors acknowledge the support ESO and its Studentship Programme provided.
		We would also like to thank Tom Robitaille, Ke Wang and Sarolta Zahorecz for the stimulating and helpful discussions.
	\end{acknowledgements}

	\bibliographystyle{aa} 
	\bibliography{ref}

	\appendix
	
	\section{Direction probability density functions of radiation fields}\label{app_mu}

		As mentioned in Sect.\ \ref{methods} the isotropy of the radiation field that heats up the dust in our simulations strongly depends on the directions of the launched photon packages.
		In the case of a point source, no direction is preferred.
		Mathematically, that means that the probability function, $p(\mu)$ with $\mu$ = cos($\theta$), is completely independent of $\mu$.
		The easiest way to implement this is by defining
		\begin{equation}
			p_{\rm point}(\mu_{\rm point}) = 1 \, . \label{prob_point}
		\end{equation}
		In the Monte Carlo (MC) code the directions are normally set such that the PDF which is given by
		\begin{equation}
			PDF(\mu_{\rm point}) = \frac{ \int_{\mu_{\rm min}}^{\mu_{\rm point}} p(\mu') d\mu' }{ \int_{\mu_{\rm min}}^{\mu_{\rm max}} p(\mu') d\mu' } \stackrel{!}{=} \xi \, . \label{def_pdf}
		\end{equation}
		This needs to equal a random number $\xi$, which is set for each photon package individually.
		We note that, although the function is similar, the PDFs in this section describe the probability of directions into which the photon packages are launched, whereas the PDFs in the main part of this paper illustrate the probability of finding a certain value of total column density or dust temperature in the corresponding maps.

		For a point source, the PDF is then
		\begin{equation}
			PDF(\mu_{\rm point}) = \frac{ \int_{-1}^{\mu_{\rm point}} 1 d\mu' }{ \int_{-1}^{1} 1 d\mu' } = \frac{\mu_{\rm point} + 1}{2} \stackrel{!}{=} \xi \label{pdf_point}
		\end{equation}
		or in terms of $\mu$, which is used to calculate the direction vector within MC,
		\begin{equation}
			\mu_{\rm point} = -1 + 2 \cdot \xi \, . \label{mu_point}
		\end{equation}
		Eq.\ (\ref{mu_point}) is not only used when new photon packages are launched from a point source, but also when photon packages are re-emitted.

		Our external radiation field, as is described in Sect.\ \ref{methods}, is supposed to be isotropic. 
		This means that the observer needs to see the same total flux independently from the viewing angle.
		We can only realise this if we vary the probability of photon packages being ejected into certain direction with $\mu$.

		This becomes understandable if we consider the following cases.
		If $d\Omega$ is the solid angle within, the observer sees the area $dA$ on a plane.
		If the line of sight is perpendicular to the plane, $\mu_{\rm iso}$ = 1, $dA$ contains a smaller fraction of the plane than if the line of sight becomes parallel to the plane, $\mu_{\rm iso} \, \approx$ 0.
		Since the total flux within $d\Omega$ needs to be constant, the flux density on the plane needs to decrease with decreasing $\mu$. 
		We can easily achieve that by setting
		\begin{equation}
			p_{\rm iso}(\mu_{\rm iso}) = \mu_{\rm iso} \, . \label{prob_iso}
		\end{equation}
		We note that contrary to the point source we want the photon packages to be launched only in the forwards direction since we want the photon packages to move from the edges into the box instead of leaving it directly, without any interacting with the dust.
		By this, the minimal angle $\mu_{\rm min}$ shifts from -1 to 0.
		Consequently, the PDF is then 
		\begin{equation}
			PDF(\mu_{\rm iso}) = \frac{ \int_{0}^{\mu_{\rm iso}} \mu' d\mu' }{ \int_{0}^{1} \mu' d\mu' } = \mu_{\rm iso}^2 \stackrel{!}{=} \xi  \, ,\label{pdf_iso}
		\end{equation}
		which relates to
		\begin{equation}
			\mu_{\rm iso} = \sqrt{\xi} \, . \label{mu_iso}
		\end{equation}

		These considerations and resolutions are consistent with those presented by \citet{Lomax2016}.
	
	\section{Benchmarks}\label{app_bench}
		\begin{figure*}[htb]
			\includegraphics[width=\textwidth]{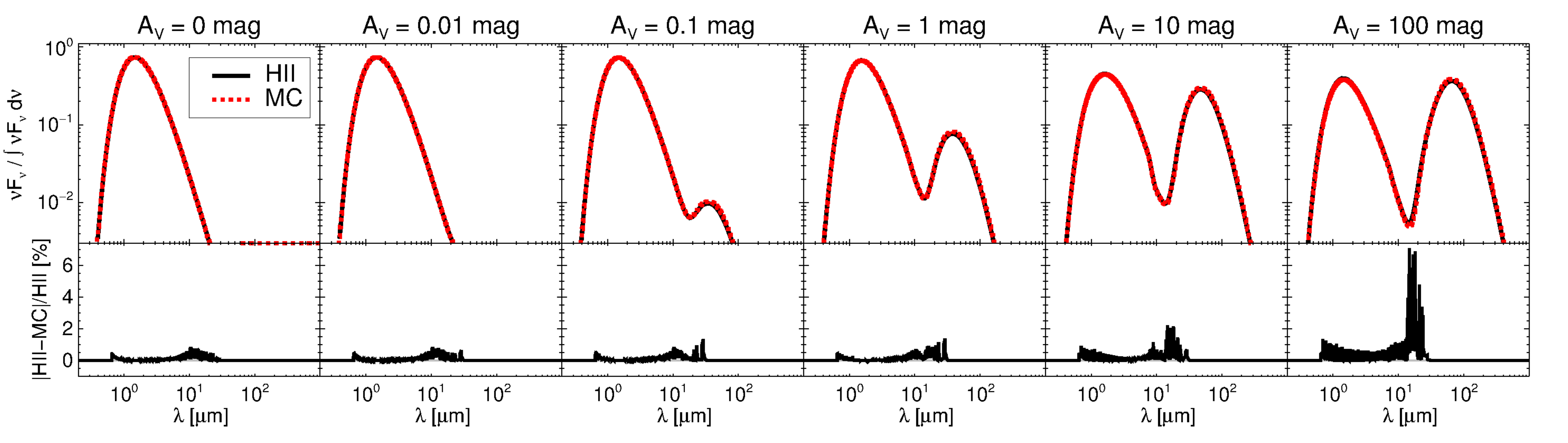}
			\caption{Results of the benchmark tests.
				\textit{Top panel:} SEDs of dust spheres that are heated externally by a diluted black body for various extinctions (as labelled).
				The SEDs are normalised to the total energy.
				Results are shown for HII (black solid) and MC (magenta dashed).
				\textit{Bottom panel:} Residuals are given as absolute relative errors (\%) of MC compared to HII.
				We see that the results derived by both methods agree well with each other.
				}
			\label{pic_bench_av}
		\end{figure*}

		We confirm the correctness of the new implementations within our MC code that we have described in Sect.\ \ref{methods} and used in \linebreak Sect.\ \ref{fila} by benchmarking it against a one-dimensional RT code called HII \citep{Kruegel2003}.
		HII solves the radiative transfer equations for dusty spheres.
		Thus, we use a homogeneous sphere for the benchmark runs.
		The sphere has an inner, $R_{\rm in}$, and outer radius, $R_{\rm out}$.
		The volume within $R_{\rm in}$ is empty, as is the volume outside $R_{\rm out}$.
		The dust is distributed between $R_{\rm in}$ and $R_{\rm out}$ and has a constant density, $\rho$.
		We subdivide the total volume into $n_{\rm x} \, \times \, n_{\rm y} \, \times \, n_{\rm z}$ cubes in x-, y-, and z- direction, respectively, having the physical edge length, $d_{\rm GW}$.
		The enclosed dust mass, $M_{\rm d}$, is given by
		\begin{equation}
			M_{\rm d} =\frac{4 \, \pi}{3} \rho  \left( R_{\rm out}^3 - R_{\rm in}^3 \right) \, , \label{equ_def_mass}
		\end{equation}
		and the total extinction $A_{\rm V}$ by
		\begin{equation}
			A_{\rm V} = \int^{R_{\rm out}}_{R_{\rm in}} \rho(r) \, \kappa_{\rm V} \, dr = \rho \cdot \kappa_{\rm V} \cdot \left( R_{\rm out} - R_{\rm in} \right) \, , \label{equ_def_av}
		\end{equation}
		where $\kappa_{\rm V}$ is the total dust opacity in the V band \linebreak ($\lambda_{\rm V}$ = 0.55 $\mu$m).
		Table \ref{tab_bench_setup} summarises the values we use for the benchmark sphere.

		We use the same composition of dust species as presented in Sect.\ \ref{fila}, but choose abundances of $\chi_{\rm aC}$ = 2.0 $\times$ 10$^{-4}$ and \linebreak $\chi_{\rm Si}$ = 3.1 $\times$ 10$^{-5}$ relative to H$_2$ for the benchmark runs.

		HII uses a one-dimensional radial grid.
		It is set in a way that the difference of the local optical depth in the V band, $\tau_V$, between two grid points is smaller than 0.1.
		Thus, the distance between two grid points depends on the density profile within the sphere, as does the required number of grid points.

		Both codes heat the dust by an external, isotropic radiation field.
		For simplicity we use a black body that is diluted by the factor $w$.
		
		\begin{table}
			\begin{center}
			\begin{tabular}{l|c|c}
				Quantity & Symbol & Value \\ \hline

				Inner radius & $R_{\rm in}$ & 1.0 $\times$ 10$^{14}$ cm \\
				Outer radius & $R_{\rm out}$ & 3.4 $\times$ 10$^{15}$ cm \\
				Edge length of major grids & $d_{\rm GW}$ & 1.0 $\times$ 10$^{14}$ cm \\
				Major grids in x/y/z direction & $n_{\rm x,y,z}$ & 35 \\
				Total luminosity & $L_{\rm tot}$ & 1.0 L$\odot$ \\
				Black body temperature & $T_{\rm BB}$ & 2,500 K \\
				Dilution factor & $w$ & 5.9 $\times$ 10$^{-9}$
			\end{tabular}
			\end{center}
			\caption{Summary of parameters used for benchmark models.
				We use the given numbers for defining the spherical volumes, as well as the radiation field for both codes. 
				The radiation field is given by a diluted black body.
			}
			\label{tab_bench_setup}
		\end{table}

		Our tests intend to verify that our modifications within the MC code agree with HII, independently of the optical thickness.
		Our sample contains six test cases:
		\begin{itemize}
			\item An empty sphere ($A_{\rm V}$ = 0 mag), \\
				(this model verifies that the photon package propagation process works correctly by reproducing our input radiation field)
			\item Three (very) optically thin models ($A_{\rm V}$ = 0.01, 0.1, 1 mag), and
			\item Two models with optically thick dust ($A_{\rm V}$ = 10, 100 mag).
		\end{itemize}
		\vspace*{-0.3\baselineskip}
		Table \ref{tab_bench_av} offers more details on the models.

		\begin{table}
			\begin{center}
			\begin{tabular}{c|c|c}
				Total extinction & Dust density & Dust mass \\ \hline
				$A_{\rm V}$ [mag] & $\rho$ [g cm$^{-3}$] & $M_{\rm d}$ [M$_\odot$] \\ \hline
				0  & 0.0 & 0.0 \\
				$10^{-2}$ & $10^{-22}$ & 8.53 $\times$ 10$^{-9}$ \\
				$10^{-1}$ & $10^{-21}$ & 8.53 $\times$ 10$^{-8}$ \\
				$10^{0}$  & $10^{-20}$ & 8.53 $\times$ 10$^{-7}$ \\
				$10^{1}$  & $10^{-19}$ & 8.53 $\times$ 10$^{-6}$ \\
				$10^{2}$  & $10^{-18}$ & 8.53 $\times$ 10$^{-5}$ \\
			\end{tabular}
			\end{center}
			\caption{Summary of parameters used for setting-up the benchmark.
				We list the total extinction, $A_{\rm V}$, of the dust along the line of sight through the spheres, the dust density, $\rho$, and the dust mass, $M_{\rm d}$.
			}
			\label{tab_bench_av}
		\end{table}

		Fig.\ \ref{pic_bench_av} shows the results of our runs.
		In the top panel we see the normalised SEDs computed by HII and MC. 
		In the case of MC, we derive the SEDs by counting photons.
		That means that MC counts the number of photon packages for each frequency bin by checking the final frequency of these photon packages when they are about to leave the data cube towards the observer.
		Since each photon package and frequency bin contains the same amount of energy, these numbers are easily converted into fluxes.
		As \citet{Heymann2012} have discussed, it is essential to use a large number of photon packages to reduce the Poisson noise in MC.
		We use 8 192 frequency bins, each containing 10$^5$ photon packages for our external radiation field, which is sufficient for a good signal-to-noise ratio.

		In most cases, the SEDs are identical within $\leq$ 5\% in relative errors.
		However, there are some parts where the differences are higher (especially the $A_{\rm V}$ = 100 mag model).
		The differences are caused by two effects:
		(a) HII and MC use different frequency grids. 
		MC requires that all photon packages within the individual frequency bins contain the same energy.
		HII does not need such an equipartition in energy and is free to bin the frequency range arbitrarily.
		Thus, when comparing the models, we have to use grid points that are only close-by, but not necessarily identical.
		This approach introduces a certain level of error, especially in regions where the SED is steep.
		(b) The spatial grids are not identical which influences the temperature profiles.
		At higher extinctions it is important to have a good resolution at the outer edges of the sphere since the temperature gradient is maximal here.
		Having a low resolution at this point means that the temperature variations are averaged out and the dust is colder with a grid cube than it is supposed to.
		HII treats that problem by automatically binning the grid accordingly.
		In MC the grid is not refined automatically.
		Thus, the user needs to be aware of this problem when setting the grid up.

		Both issues can be fixed by, for example, using more and smaller grid cells.
		This increases the spatial resolution, but also increases the computational time significantly.
		With our set-ups the residuals are sufficiently small for our further investigations.

	\onecolumn
	\thispagestyle{headings}
	\section{Additional Images}\label{app_images}

		\begin{figure}[h]
			\centering
			\includegraphics[width=0.81\textwidth]{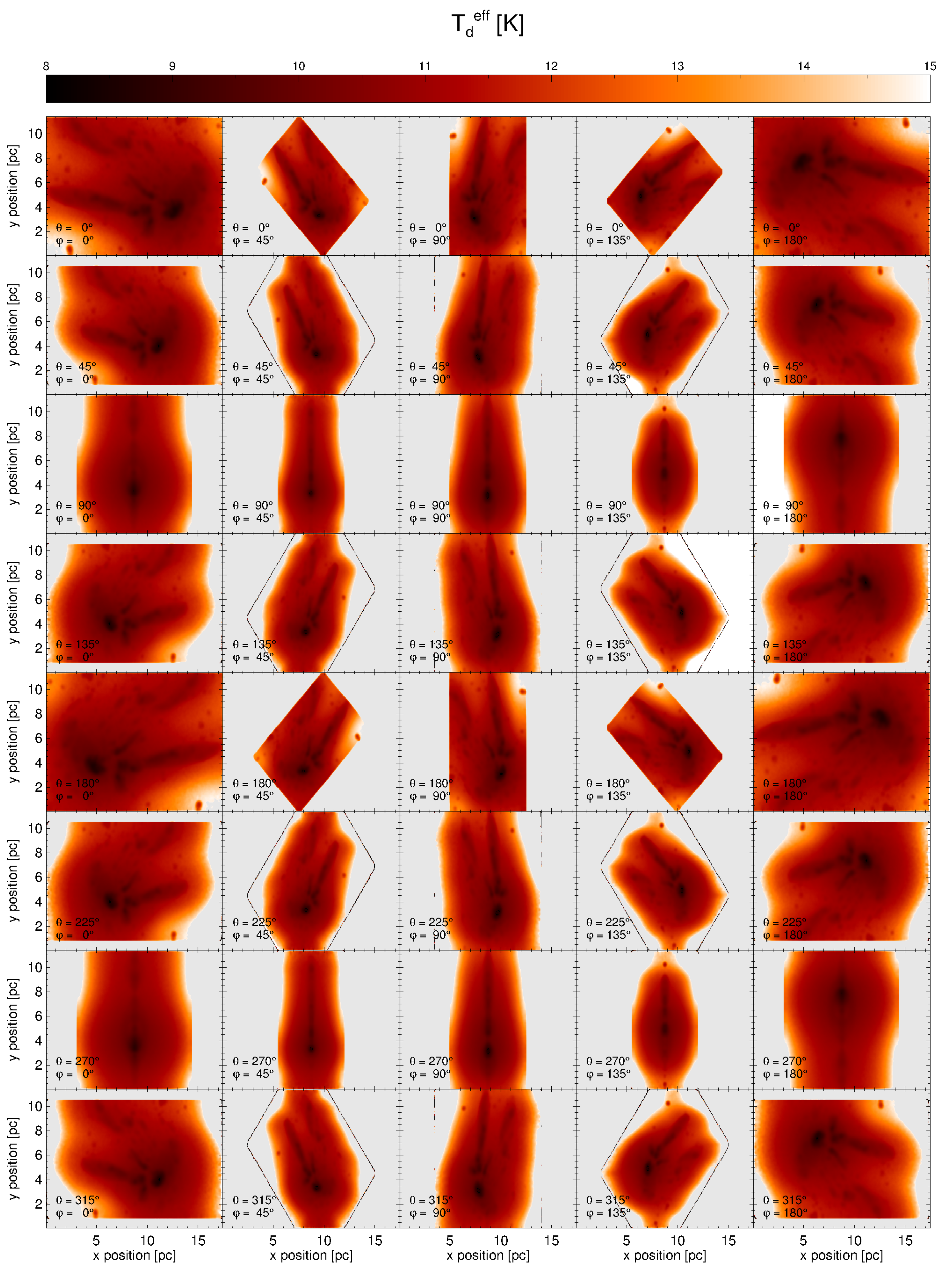}
			\caption{$\rho$ Ophiuchi cloud model. Maps of effective dust temperature along the line of sight, $T_{\rm d}^{\rm eff}$, that we derive by pixel-by-pixel SED fitting (see Sect.\ \ref{fila}). }
			\label{pic_rhooph_tmap}
		\end{figure}

		\begin{figure}
			\centering
			\includegraphics[width=0.81\textwidth]{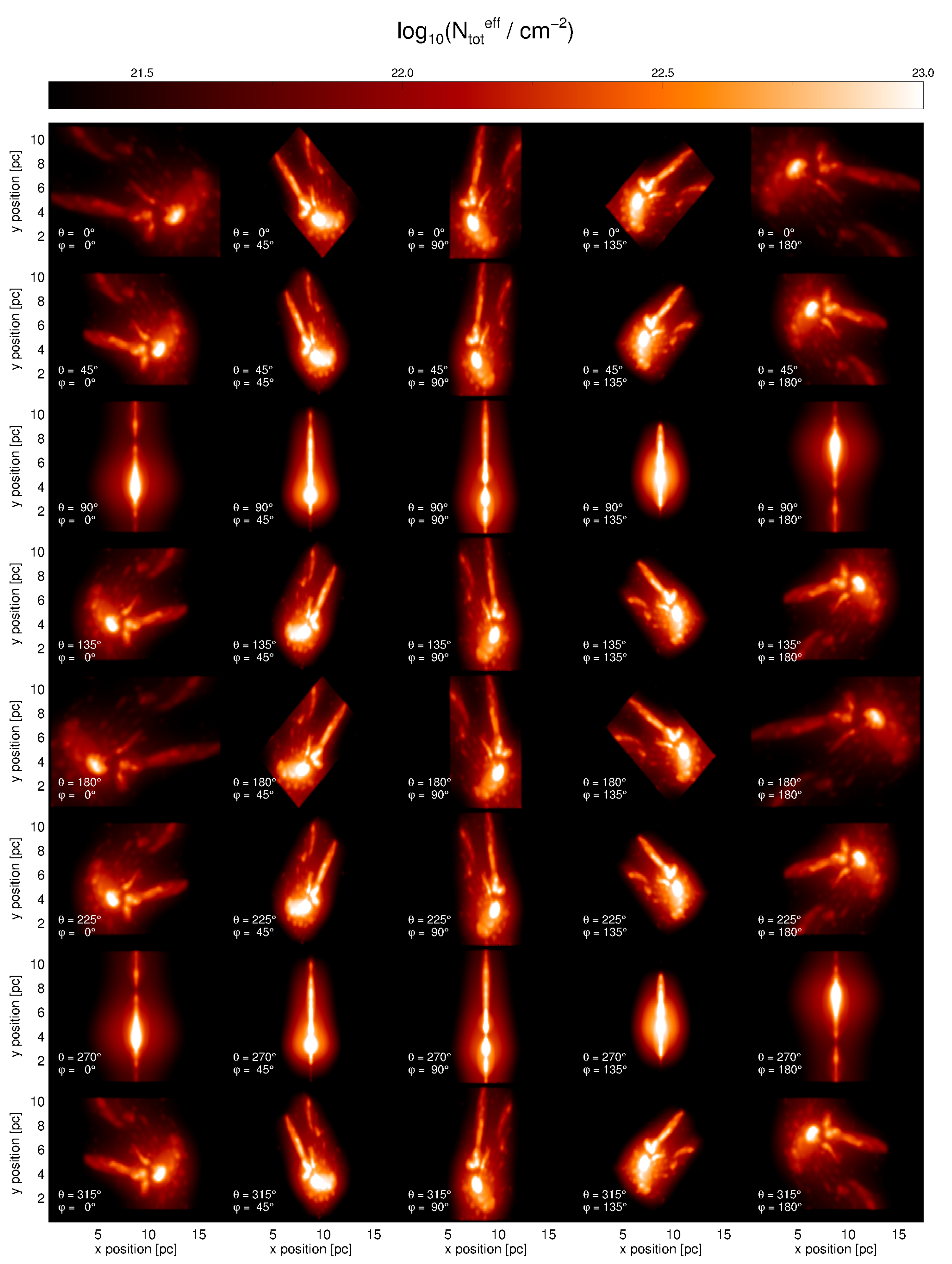}
			\caption{$\rho$ Ophiuchi cloud model. Like Fig.\ \ref{pic_rhooph_tmap} showing the effective total column density along the line of sight, $N_{\rm tot}^{\rm eff}$.}
			\label{pic_rhooph_nmap}
		\end{figure}

		\begin{mysidewaysfigure}
			\includegraphics[width=0.95\textwidth]{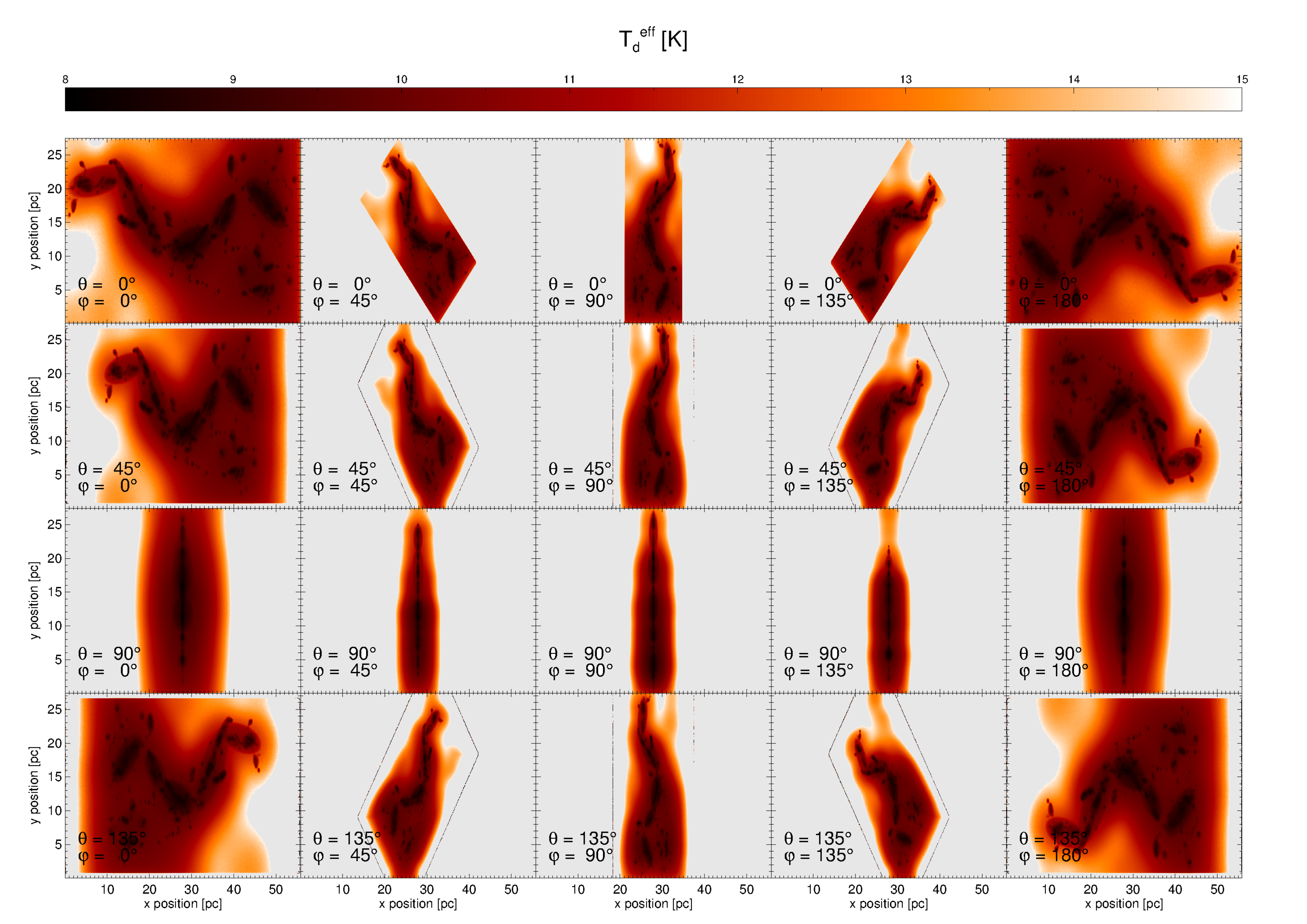}
			\caption{G11.11 \textit{Snake} model. Maps of effective dust temperature along the line of sight, $T_{\rm d}^{\rm eff}$, that we derive by pixel-by-pixel SED fitting.
			Unlike Fig.\ \ref{pic_rhooph_tmap} we do not show the maps at $\theta \, \geq$ 180$^\circ$ due to the reasons discussed in Sect.\ \ref{apply_rhooph}.
			}
			\label{pic_g11_tmap}
		\end{mysidewaysfigure}

		\begin{mysidewaysfigure}
			\includegraphics[width=0.95\textwidth]{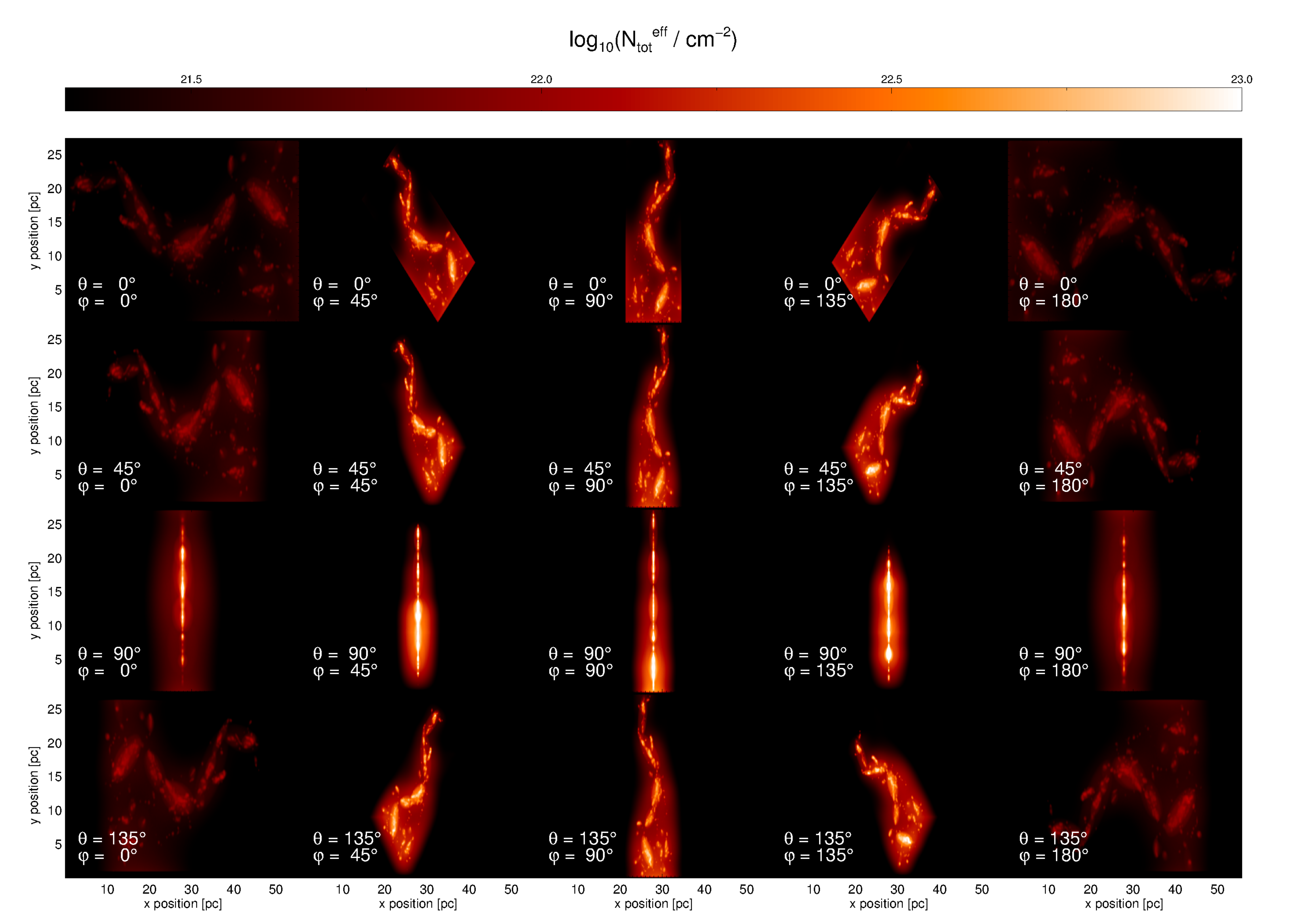}
			\caption{G11.11 \textit{Snake} model. Like Fig.\ \ref{pic_g11_tmap} showing the effective total column density along the line of sight, $N_{\rm tot}^{\rm eff}$.}
			\label{pic_g11_nmap}
		\end{mysidewaysfigure}

\end{document}